\documentclass[aps,superscriptaddress,showpacs,floatfix]{revtex4-1}

\usepackage{hyperref}
\usepackage{color}
\usepackage{graphicx}	
\graphicspath{%
  {./},%
}

\usepackage{amstext}
\usepackage{array}

\usepackage{xspace}	

\begin{document}

\title{Lattice Boltzmann simulations of a two-dimensional Fermi gas at unitarity}

\author{Jasmine Brewer} \affiliation{University of Colorado at Boulder}
\author{Miller Mendoza}\affiliation{ ETH
  Z\"urich, Computational Physics for Engineering Materials, Institute
  for Building Materials, Schafmattstrasse 6, HIF, CH-8093 Z\"urich
  (Switzerland)}
\author{Ryan E. Young$^{1}$}
\author{Paul Romatschke$^{1}$}
\date{\today}

\begin{abstract}
We present fully nonlinear dissipative fluid dynamics simulations of a trapped two-dimensional Fermi gas at unitarity using a Lattice Boltzmann algorithm. We are able to simulate non-harmonic trapping potentials, temperature-dependent viscosities as well as a discretized version of the ballistic (non-interacting) behavior. Our approach lends itself to direct comparison with experimental data, opening up the possibility of a precision determination of transport coefficients in the unitary Fermi gas. Furthermore, we predict the presence of a non-hydrodynamic component in the quadrupole mode, which should be observable experimentally.
\end{abstract}

\maketitle

\section{Introduction}

Cold Fermi gases at unitarity are examples of so-called 'Strongly Interacting Quantum Fluids' (SIQFs), which share the common characteristic that they flow around obstacles almost without friction. Other experimental examples of SIQFs seemingly include high temperature superconductors \cite{Rameau:2014gma} and hot quark gluon plasmas generated in ultrarelativistic heavy-ion collisions \cite{Luzum:2008cw,Heinz:2013th}, while the cleanest theoretical example seems to be given by black holes \cite{Policastro:2001yc}.

The last decade has seen an unprecedented development in precision experiments in these ultracold quantum gases at unitarity, both in two and three dimensions, i.e.~\cite{2008PhRvL.101m5301P,2008PhRvA..78e3609R,Cao:2010wa,2010NatPh...6..569G,2012Sci...335..563K,Vogt:2011np,2014NatPh..10..116M}. So far, theoretical descriptions of the collective behavior observed in these experiments has mostly been based on (analytic) linear or scaling solutions to hydrodynamic or kinetic equations, see e.g. Refs.~\cite{Schafer:2007pr,Schaefer:2009px,2013PhRvA..87d3612B}. While these solutions have the advantage of being fully analytical, they also have the drawback of applying only to specific, idealized situations, such as harmonic trapping potentials, fixed temperature dependence of transport coefficients and hydrodynamic behavior in the cloud's low density corona.

In the present work we follow a completely different approach in that we intend to set up numerical `experiments' that are capable of simulating the behavior of cold atomic gases for setups that include, but are not limited to, the experimentally realized scenarios. We expect that by being able to simulate realistic (e.g. non-idealized) configurations as in experiment our method could lead to much more precise extraction of transport properties from experimental data than using analytic solutions. Moreover, by being able to simulate configurations that are not (yet) experimentally realized we can hope to point out and possibly predict interesting cloud behavior. This work is meant to be a first step in this direction, in that we limit ourselves to discuss only the case of the collective modes of a trapped, two-dimensional Fermi gas at unitarity and ignoring heat 
conduction effects. Generalizations of our approach to three dimensions, Bose gases and inclusion of heat conduction are straightforward and will be considered in future work. 

The simulations we perform are meant to describe the bulk evolution and transport phenomena occurring in an atomic cloud or clouds, in particular ignoring the effect of quantum transitions and quantum tunneling between spatially separated clouds, and outside the superfluid regime. In the absence of dissipation, such dynamics would be well described by one-component hydrodynamics. Given that we are interested in the effects of dissipation (transport), and that dissipation is always important in the low density corona of the cloud or clouds, a description in terms of the so-called Lattice Boltzmann (LB) algorithm \cite{Succibook} may be well suited to offer a realistic description of the atomic cloud dynamics. Despite their name and origin, it should be pointed out that Lattice Boltzmann simulations are not limited to situations with well-defined quasi-particles or weak coupling situations, which is why they are well suited for simulating SIQFs.

Compared to recent results in the literature, our study is closely related to Ref.~\cite{Pantel:2014jfa}, where the Boltzmann equation was solved numerically using the test particle method. Compared to Ref.~\cite{Pantel:2014jfa}, our method has the disadvantage of not accurately describing the non-interacting regime of the cold atom gas on a quantitative level. However, the present approach has the advantage of allowing for arbitrary equations of state, the straightforward simulation of shock waves as well as being considerably cheaper in terms of computational cost. 

Furthermore, the present work is related to Ref.~\cite{Bluhm:2015raa}, where non-hydrodynamic elements from the Boltzmann equation were used to improve the hydrodynamic evolution equations in the non-interacting regime. Similar to Ref.~\cite{Pantel:2014jfa}, the algorithm used in Ref.~\cite{Bluhm:2015raa} is superior to our approach in that the non-interacting regime is treated exactly, but unlike our setup does not allow for non-ideal equations of state.

Moreover, our work is similar to Refs.~\cite{0295-5075-97-1-16003,PhysRevA.85.045601} where the Boltzmann equation was solved numerically in the relaxation time approximation. Compared to our work, the method in Refs.~\cite{0295-5075-97-1-16003,PhysRevA.85.045601} can obtain reliable results also in the non-interacting regime, but our approach is computationally cheaper.

Finally, our work is closely related to Refs.~\cite{2013PhRvA..88e3616C,2014JPhCS.497a2028U} where the Boltzmann equation was studied in the moment approximation. This is because the lattice Boltzmann method is essentially a moment approximation to the Boltzmann equation packaged in a computationally highly efficient form. However, compared to the moment approximation of the Boltzmann equation, the lattice Boltzmann framework offers more flexibility by allowing for non-ideal equations of state as well as interaction terms that are not realizable in a particle picture. In this sense, the lattice Boltzmann framework is closer to hydrodynamics than to the actual Boltzmann equation, being an effective theory for low frequency, small wave-number transport.

This work is organized as follows: In section \ref{sec:boltz} we give a derivation of the Lattice Boltzmann framework for the use of simulating cold trapped quantum gases at unitarity. Section \ref{sec:results} contains our results for trapped two-dimensional gases with ideal and non-ideal equations of state in harmonic and Gaussian traps. We present our conclusions in section \ref{sec:conc} and include details about our numerical scheme in an appendix.

\section{The Lattice Boltzmann Framework}
\label{sec:boltz}

\subsection{Kinetic Description of Trapped Cloud}
The Boltzmann equation for a single particle distribution function $f(t,{\bf x},{\bf v})$ is given by
\begin{equation}
\label{eq:Boltz}
\left[\partial_t+{\bf v}\cdot \nabla-\frac{1}{m}\nabla U(x)\cdot \nabla^{(v)}\right] f={\cal C}[f]
\end{equation}
where ${\bf v},m$ are the particle's velocity and mass, $U(x)$ is the trapping potential and ${\cal C}[f]$ is the collision term that depends on the particle interactions. From the single particle distribution function one can define the local mass density $\rho$, macroscopic (fluid) velocity ${\bf u}$ and total (kinetic plus internal) energy density $\epsilon$ through the following integral moments:
\begin{equation}
\rho\equiv m \int d^{D}{\bf v} f\,,\quad
\rho {\bf u}\equiv m \int d^{D}{\bf v} {\bf v}f\,,\quad
\epsilon \equiv \frac{m}{2} \int d^{D}{\bf v} {\bf v}^2 f \,,
\label{eq:therm}
\end{equation}
where $D$ is the number of space dimensions.

It is well known that if the interactions are strong enough, then the dynamical evolution of the macroscopic quantities $\rho,{\bf u},\epsilon$ will obey the equations of (Navier-Stokes) hydrodynamics. However, in this limit the microscopic details of the particle interactions become unimportant, and the hydrodynamic evolution for $\rho,{\bf u},\epsilon$ stay {\it unchanged} if one replaces the (complicated) collision term with a simple BGK-type ansatz as long as this ansatz obeys the conservation of mass, momentum and energy:
\begin{equation}
\label{eq:collterm}
{\cal C}[f]=-\frac{f-f_{\rm eq}}{\tau_R}\,,
\end{equation}
where $f_{\rm eq}$ is the local equilibrium distribution function and $\tau_R$ is the (local) relaxation time. It is straightforward to find static equilibrium solutions to Eq.~(\ref{eq:Boltz}), which in units where $c=k_b=\hbar=1$ can be expressed in terms of the macroscopic variables as
\begin{equation}
\label{eq:eqdist1}
f_{\rm eq}(t,{\bf x},{\bf v})=\frac{\rho e^{\frac{-({\bf v-u})^2}{2 c_s^2(T)}}}
{m c_s^D(T) \pi^{D/2}}\times\left\{\begin{array}{c}1\,,D=2\\ \frac{1}{\sqrt{8}}\,, D=3\end{array}\right.\,,
\end{equation}
with $c_s^2(T)=\frac{T}{m}$ the local speed of sound squared and a proportionality constant that depends on the number of space dimensions $D$. Note that this implies  $\epsilon=\frac{1}{2}\rho {\bf u}^2+\frac{D}{2}\rho c_s^2(T)$. The careful reader will at this point worry about the choice of a Boltzmann-type distribution function (\ref{eq:eqdist1}), which may seem a bad approximation for describing quantum gases where actual particle distribution functions should be given by Fermi-Dirac or Bose-Einstein statistics. There are two important points to consider for this issue. One, our description is aimed at the evolution of the macroscopic system variables and we do not aim for a correct description of the actual particle distribution. Two, for the bulk evolution of the system the information about the actual particle distribution enters only through the equation of state (e.g. the relation between pressure and density), and we will demonstrate how to correctly implement this information in the following section.

As a consequence of the unimportance of the microscopic collision term for the hydrodynamic evolution of the system, the original restriction of Eq.~(\ref{eq:Boltz}) to well-separated particle degrees of freedom can be lifted, since Eqns.~(\ref{eq:Boltz}),(\ref{eq:collterm}) no longer explicitly refer to particles anymore.
Thus, Eqns.~(\ref{eq:Boltz}),(\ref{eq:collterm}) can be viewed as a system of effective equations describing the dynamic evolution of the macroscopic system variables $\rho,{\bf u},\epsilon$ in the limit of strong interactions ($\tau_R\rightarrow 0$). In particular, this implies that Eqns.~(\ref{eq:Boltz}),(\ref{eq:collterm}) can give an accurate effective system description even in situations where the original Boltzmann equation (\ref{eq:Boltz}) may no longer be well-defined.
Conversely, in the case of weak interactions (in particular in the ballistic regime $\tau_R\rightarrow \infty$), the ansatz (\ref{eq:collterm}), while qualitatively correct, may offer only a poor quantitative approximation to the exact microscopic collision term ${\cal C}[f]$. These considerations set the regime of applicability for the Lattice Boltzmann framework outlined below.

Let us now rescale coordinates so that we work in dimensionless units adapted to the cold atoms system characterized by some transverse size $R_\perp$ and some frequency $\omega_\perp$:
\begin{equation}
\label{coords}
{\bf x}={\bf \bar{x}} R_\perp\,,\quad t=\bar{t}/\omega_\perp\,, \quad {\bf v}={\bf \bar{v}} R_\perp \omega_\perp\,,\quad {\bf u}={\bf \bar{u}} R_\perp \omega_\perp\,,
\end{equation}
so that (\ref{eq:Boltz}) becomes
$$
\left[\partial_{\bar{t}}+{\bf \bar{v}}\cdot \bar{\nabla}-\frac{1}{R_\perp^2 \omega_\perp^2 m}\bar{\nabla} U(\bar{x})\cdot \nabla^{(\bar{v})}\right] f=-\frac{f-f_{\rm eq}}{\tau_R \omega_\perp}\,.
$$

In the following, the initial condition considered is that for a static cloud in equilibrium, which to a good approximation will be isothermal with an initial temperature $T_0$.
%
Introducing in addition a calculational (constant) parameter $c_L$, which will be referred to as the 'lattice speed' below, we chose our unit system parameter $R_\perp$ to be given as
\begin{equation}
\label{eq:setR}
R_\perp=\sqrt{\frac{T_0}{m c_L^2  \omega_\perp^2}}\
\end{equation}
such that 
\begin{equation}
\left[\partial_{\bar{t}}+{\bf \bar{v}}\cdot \bar{\nabla}-\frac{c_L^2}{T_0}\bar{\nabla} U(\bar{x})\cdot \nabla^{(\bar{v})}\right] f=-\frac{f-f_{\rm eq}}{\tau_R \omega_\perp}\,,
\label{eq:resBoltz}
\end{equation}
with $\theta=\frac{T}{T_0}$. Note that in these units, for a cloud that is at rest with temperature $T_0$ and trapping potential $U$, the mass density initially is given by 
\begin{equation}
\label{eq:rho}
\rho\propto \exp{\left(-\frac{U({\bar x})}{T_0}\right)}\,.
\end{equation}

\subsection{From Boltzmann to Lattice Boltzmann}
\label{sec:lbone}

So far we have discussed a treatment of trapped atomic gases in continuum kinetic theory. However, it turns out that as long as we are only concerned with the evolution of the macroscopic variables (\ref{eq:therm}) we do not actually need to keep the full continuum information of $f(t,{\bf x},{\bf v})$ in velocity space. Specifically, since (\ref{eq:therm}) refer only to a finite number of moments of the distribution function, we can obtain an exact representation of these integrals by expanding $f$ in a series of polynomials which are orthogonal and complete with respect to some reference distribution function (see e.g. \cite{Succibook} for details of the Lattice Boltzmann method). In particular, choosing $e^{-\frac{{\bf v}^2}{2 c_L^2}}$ as a reference distribution function, the orthogonal polynomials are just the well-known Hermite polynomials. In compact index-notation, we collect these orthogonal polynomials into tensors $P_n^{i_1 i_2\ldots i_n}({\bf \bar v})$ where $n$ is the order of the polynomial and the tensorial indices $i_1,i_2,\ldots$ run from 1 to the number of space dimensions. For instance, we have
\begin{equation}
P_0({\bar v})=1\,,\quad 
P_1^i({\bar v})={\bar v}^i\,,\quad 
P_2^{ij}({\bar v})={\bar v}^i{\bar v}^j-c_L^2 \delta^{ij}\,.\quad 
\end{equation}
Because the basis functions are polynomials, one can use quadrature-type rules to exactly represent the integrals (\ref{eq:therm}). The nodes of polynomials select specific, optimized values for the values of the velocity vectors ${\bf \bar v}$, such that one is left with a velocity lattice rather than a continuous collection of velocities. Labeling the individual velocity vectors of the lattice through the label $s$ (running from $1$ to $Q$, the total number of velocity vectors in the lattice), the relevant moments become
\begin{eqnarray}
\frac{\rho}{m R_\perp^D \omega_\perp^D}\equiv n &=& \sum_{s=1}^Q w_s f_s(t,{\bf x})\,,\nonumber\\
\frac{\rho {\bf \bar u}}{m R_\perp^D \omega_\perp^D} = n {\bf \bar u} &=& \sum_{s=1}^Q w_s {\bf \bar v}_s f_s(t,{\bf x})\,,\nonumber\\
\frac{\epsilon}{\frac{m}{2} R_\perp^{D+2} \omega_\perp^{D+2}}=n{\bf \bar u}^2+n c_L^2 \frac{P(n,T)}{n T_0}D &=& \sum_{s=1}^Q w_s {\bf \bar v}_s^2 f_s(t,{\bf x})\,,
\label{eq:thermLB}
\end{eqnarray}
where $w_s$ are suitably chosen integration weights and we have introduced the pressure $P(n,T)$. It is worth stressing that while the continuum solution $f_{\rm eq}$ in (\ref{eq:resBoltz}) demands an ideal gas equation of state $P(n,T)=n T$, this restriction is (to some degree) lifted in the discretized version $f_s(t,{\bf x})$. In particular, this implies that non-ideal equations of state $P(n,T)$ can be simulated using the LB method. 

In the following we will work with a previously defined velocity lattice that is known as D2Q25 (D=2 space dimensions, and $Q=25$ velocity vectors), for which $c_L^2=1-\sqrt{\frac{2}{5}}$. The individual velocities and weights for this lattice are given in Table \ref{table:D2Q25}.

\begin{table}
\centering
\parbox{0.45\textwidth}{
\begin{tabular}{| >{$}c<{$} | >{$}c<{$} | }
\hline
\bf{v}_i & w_i \\
\hline
(0,0) & \iota_0 \\
(0,\pm 1)  \text{ and } (\pm 1,0) & \iota_0 \iota_1 \\
(\pm 1,\pm 1) & \iota_1^2 \\
(0,\pm 3) \text{ and } (\pm 3,0) & \iota_0 \iota_3 \\
(\pm 1,\pm 3) \text{ and } (\pm 3,\pm 1) & \iota_1 \iota_3 \\
(\pm 3,\pm 3) & \iota_3^2 \\
\hline
\end{tabular}
\label{table:D2Q25}
}
\qquad
\begin{minipage}[c]{0.45\linewidth}
\centering
\includegraphics[width=0.45\linewidth]{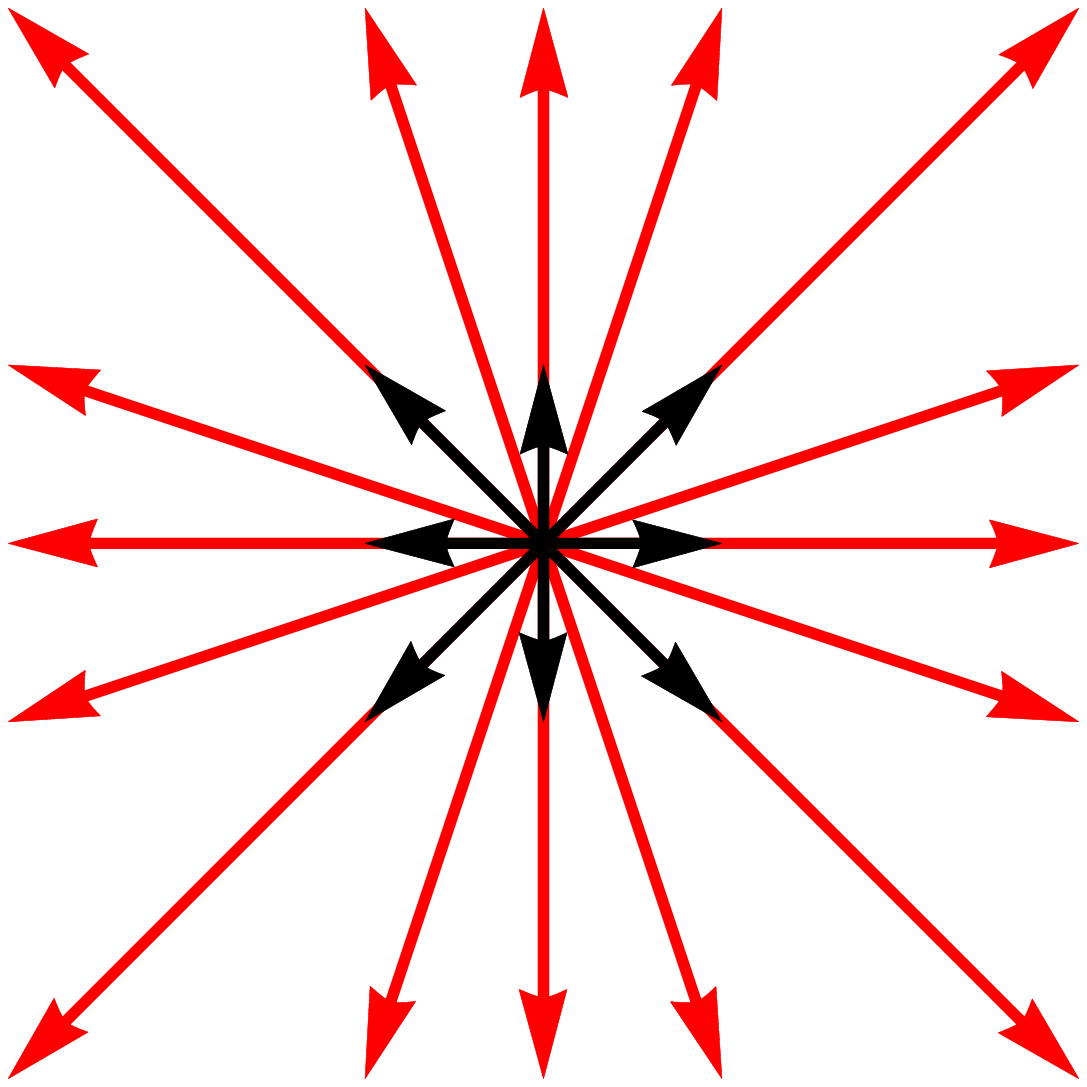}
\end{minipage}
\caption{Left: Velocities and weights of the D2Q25 lattice, where $ \iota_0 = \frac{4}{45}(4+\sqrt{10}) $, $ \iota_1 = \frac{3}{80}(8-\sqrt{10})$, $ \iota_3 = \frac{1}{720}(16-5\sqrt{10})$. Right: Illustration of the velocity vectors for the D2Q25 lattice. Note that for the D2Q25 lattice, $c_L^2=1-\sqrt{\frac{2}{5}}$. }
\end{table}

Expanding the equilibrium distribution function $f_{\rm eq}$ in terms of the reference distribution function and up to third order in polynomials (cf. Ref.~\cite{1997comp.gas.12001S}) gives
\begin{eqnarray}
\label{eq:BE}
f_{\rm eq}\propto\frac{n}{\pi^{D/2}c_s^D(T_0)} e^{-\frac{{\bf \bar{v}}^2}{2 c_L^2}}
&&\left[1+\frac{{\bf \bar u}\cdot {\bf \bar v}}{c_L^2}\left(1+\frac{P/(n T_0)-1}{2 c_L^2}({\bf \bar v}^2-(D+2) c_L^2)\right)+\frac{({\bf \bar u}\cdot {\bf \bar v})^2}{2 c_L^4}-\frac{{\bf \bar u}^2}{2 c_L^2}\right.\nonumber\\
&&\left.+\frac{P/(n T_0)-1}{2 c_L^2}\left({\bf \bar v}^2-D c_L^2\right)
+\frac{({\bf \bar u}\cdot {\bf \bar v})^3}{6 c_L^6}-\frac{{\bf \bar u}^2 ({\bf \bar u}\cdot {\bf \bar v})}{2 c_L^4}
\right]\,,
\end{eqnarray}
with a proportionality constant that is specified in Eq.~(\ref{eq:eqdist1}).
It is straightforward to verify that truncation of $f_{\rm eq}$ at this order of expansion still leads to  exact hydrodynamic evolution equations for the macroscopic quantities $\rho,{\bf u},\epsilon$ (for small gradients of the temperature). For a trapped gas, we also need an expansion of the force term in Eq.(\ref{eq:resBoltz}) in terms of orthogonal polynomials:
\begin{equation}
{\bf F}\cdot \nabla^{(\bar v)} f =  e^{-\frac{{\bf \bar{v}}^2}{2 c_L^2}} \sum_n P_n({\bf \bar v}) a_n(t,{\bf \bar x})\,,\quad {\bf F}\equiv \frac{c_L^2}{T_0}\bar\nabla U({\bar x})\,,
\end{equation}
where we have suppressed the tensorial indices for simplicity. Using the orthogonality of the polynomials, a straightforward calculation gives the coefficients $a_n$ and finally the representation of the force term as 
\begin{eqnarray}
\label{eq:force}
{\bf F}\cdot \nabla^{(\bar v)} f &=&-\frac{n}{\pi^{D/2}c_s^D(T_0)} e^{-\frac{{\bf \bar{v}}^2}{2 c_L^2}} \left[
-\frac{{\bf \bar u}\cdot {\bf F}}{c_L^2}\left(1
+\frac{{\bf \bar u}\cdot {\bf \bar v}}{c_L^2}\right)
\right.\nonumber\\
&&\left.
+\frac{{\bf \bar v}\cdot {\bf F}}{c_L^2}\left(1+
\frac{{\bf \bar u}\cdot {\bf \bar v}}{c_L^2}+
\frac{({\bf \bar u}\cdot {\bf \bar v})^2}{2 c_L^4}-\frac{{\bf \bar u}^2}{2 c_L^2}+\frac{P/(n T_0)-1}{2}\left(\frac{{\bf \bar v}^2}{c_L^2}-(D+2)\right)\right)
\right]\,.
\end{eqnarray}

With the equilibrium distribution and the force term suitably discretized, one now needs to specify the discretization of the space and time derivatives in Eq.(\ref{eq:resBoltz}). Here we use the simplest version which is to rewrite
\begin{equation}
\label{eq:SLB}
\left[\partial_{\bar t} + {\bf \bar v}\cdot \bar \nabla \right] f(\bar t,{\bf \bar x})
\simeq  \frac{f(\bar t+\delta \bar t,{\bf x+\bar v}\delta \bar t)-f(\bar t,{\bf \bar x})}{\delta \bar t}\,.
\end{equation}
Since the momentum lattice D2Q25 is space-filling, this particular choice implies that if we discretize space on a cubic lattice, at every time step 'particles' stream from one lattice site to the next lattice site, so there is no need for any interpolation schemes for Eq.~(\ref{eq:SLB}). However, note that (\ref{eq:SLB}) is only exact up to first order in derivatives. Taking into account second-order derivatives finally leads to an evolution equation of the form
\begin{equation}
\label{eq:Boltzfinal}
f(\bar t+\delta \bar t,{\bf x+\bar v}\delta \bar t)=f(\bar t,{\bf \bar x})
\left(1-\tilde \Omega\right)+f_{\rm eq}(\bar t,{\bf \bar x})\tilde\Omega  + \delta \bar t {\bf \tilde F} \cdot \nabla^{(\bar v)} f\,,
\end{equation}
with a force ${\bf \tilde F}$, relaxation term $\tilde\Omega$ as well as macroscopic variables that include numerical modifications to continuum expressions. Denoting these (numerically corrected) quantities by a tilde, one finds
\begin{eqnarray}
\label{eq:numcorr}
\tilde n = n\,,\quad {\bf \tilde u}={\bf \bar u}-\frac{\delta \bar t}{2} {\bf F}\,,\quad
\tilde \Omega^{-1}=\frac{1}{2}+\frac{\omega_\perp \tau_R}{\delta \bar t}\,,\quad
{\bf \tilde F}={\bf F}\left(1-\frac{\delta \bar t}{2}\Omega\right)\,,\quad
\tilde T = T\,.
\end{eqnarray}
Note that for the evolution of Eq.(\ref{eq:Boltzfinal}), one first calculates e.g. 
$n {\bf \bar u}=\sum_s {\bf \bar v}_s w_s f_s$, and then performs the numerical correction, before using ${\bf \tilde u}$ to construct ${\rm f}_{\rm eq}$ through
Eq.~(\ref{eq:BE}). 

Given a force ${\bf F}$ and a temperature/density dependent relaxation time $\tau_R \omega_\perp$, the simulation algorithm then can be summarized as follows:
\begin{enumerate}
\item
Select an initial condition for $f$ (e.g. from Eq.(\ref{eq:BE}))
\item
Free-stream the components of $f$ to neighboring lattice sites through
$f(\bar t+\delta \bar t,{\bf x+\bar v}\delta \bar t)=f(\bar t,{\bf \bar x})$
\item
Calculate macroscopic variables $n,{\bf \tilde u},T$ for this new configuration
\item
Calculate $f_{\rm eq},{\bf \tilde F}$ from the macroscopic variables
\item
Correct $f(\bar t+\delta \bar t)$ according to Eq.(\ref{eq:Boltzfinal})
\item
Repeat from Step 2
\end{enumerate}

For completeness, we also note that the relaxation time $\tau_R$ is controlling the simulated ratio (and temperature/density dependence) of the shear viscosity $\eta$ over the pressure $P$ of the cloud through the relation (cf. Ref.~\cite{Dusling:2011dq})
\begin{equation}
\label{eq:tauReta}
\tau_R=\eta/P\,.
\end{equation}
Note that other transport processes (e.g. spin diffusion) would have transport times that differ from $\tau_R$, and also the relation between diffusion constant and diffusion transport time would not be identical to Eq.~(\ref{eq:tauReta}). Here we limit ourselves to just momentum transport controlled by the shear viscosity.

\section{Results}
\label{sec:results}

\subsection{2D Ideal Gas in an Harmonic Trap: Analytics}
\label{sec:ana}

Solving Eq.~(\ref{eq:resBoltz}) in two dimensions for a static equilibrium solution for $f$ in an arbitrary trapping potential with $T=T_0$ and an ideal gas equation of state $P(n, T)=n T$ gives
\begin{equation}
\label{eq:f0sol}
f_0({\bf x},{\bf \bar v})=\frac{e^{-\frac{{\bf \bar v}^2}{2 c_L^2}-
\frac{U({\bar x})}{T_0}}}{\pi c_s^2(T_0)}\,.
\end{equation}
The simplest case to study is that of a symmetric harmonic trapping potential $U(\bar{x})=\frac{T_0}{c_L^2} \frac{{\bf \bar x}^2}{2}$. A simple  solution to Eq.~(\ref{eq:resBoltz}) with harmonic potential is the case of center-of-mass oscillations (sloshing mode), which can be written as
\begin{equation}
\label{eq:anaslo}
f(\bar t,{\bf x},{\bf \bar v})
=f_{0}\left({\bar x_i}-{\bar c}(t),{\bar v_i}-{\bar c}'(t)
\right)\,,\quad c(\bar t)+c''(\bar t)=0\,.
\end{equation}
A general ansatz for the sloshing mode equation of motion is 
\begin{equation}
\label{eq:slosh}
c(\bar t)=\alpha e^{-\Gamma_S\bar t}\cos(\omega_S \bar t+{\rm const})\,,
\end{equation}
and in the idealized case at hand one finds $\Gamma_S=0,\omega_S=1$ (in units of the base frequency $\omega_\perp$).

\begin{figure}[t]
\centering
\includegraphics[width=0.45\textwidth]{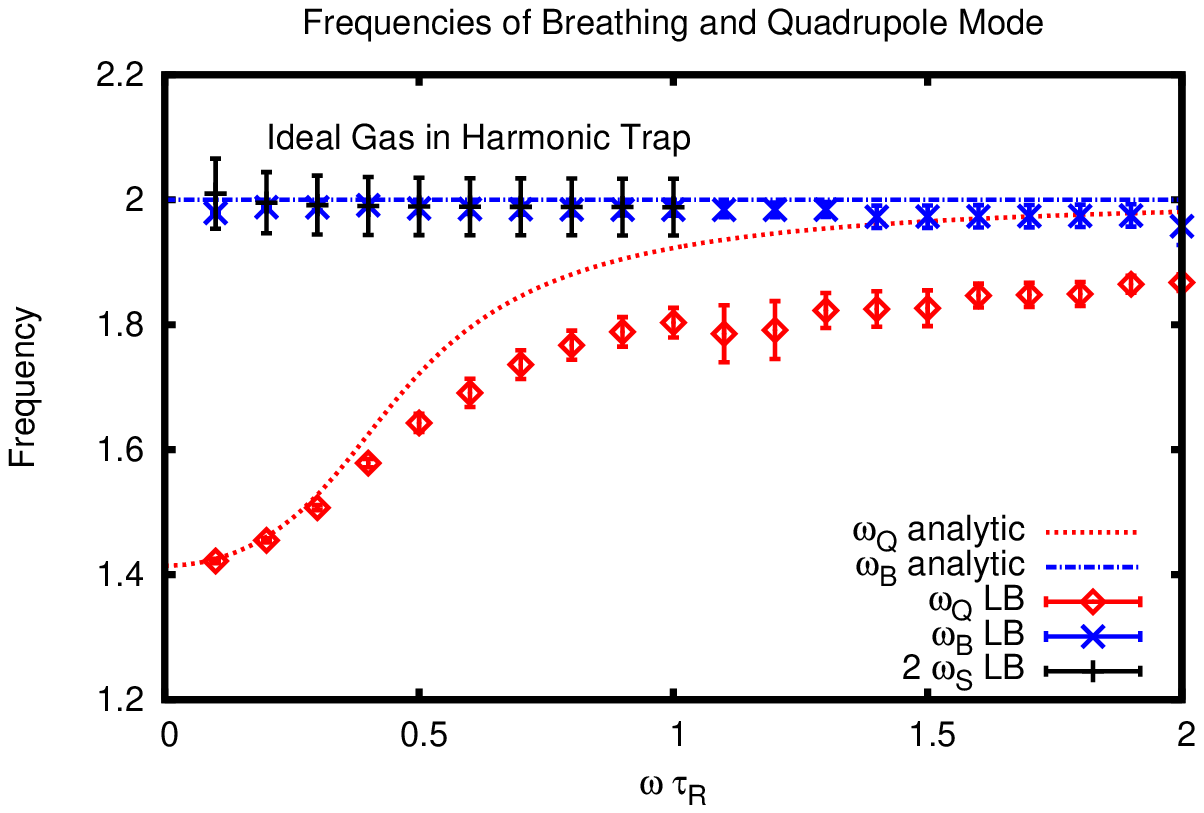}\hfill
\includegraphics[width=0.45\textwidth]{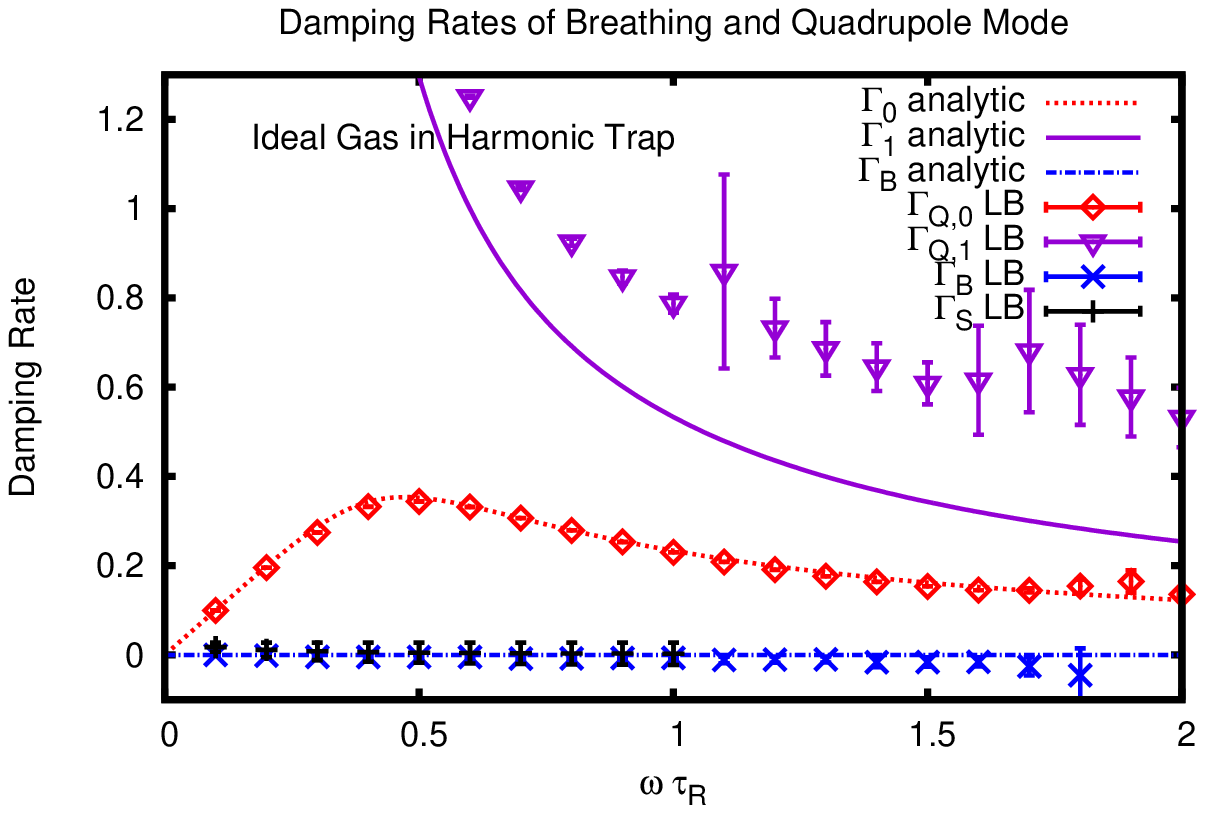}
\caption{Frequency and damping rates for the sloshing, breathing and quadrupole mode $c,\delta B,\delta Q$, respectively, of an ideal gas of atoms in an harmonic trap. Shown are exact analytic results in the limit of small amplitudes (lines) and fitted values from a fully numerical LB simulation (symbols with systematic errors from the infinite volume extrapolation). The 'hydrodynamic' regime is the limit of $\omega_\perp \tau_R\rightarrow 0$ (e.g. linear rise of $\Gamma_{Q,0}$ with $\omega_\perp \tau_R$), whereas the 'ballistic' (or 'free streaming') regime corresponds to $\omega_\perp \tau_R\rightarrow \infty$. The plot suggests a turnover point from the hydrodynamic regime to the ballistic regime at around $\omega_\perp \tau_R\sim 0.5$. The comparison between analytic and numerical LB results indicates that the fully numerical simulation reproduces the analytic results rather well even for values of the relaxation time that are approaching the ballistic (free streaming) limit. The only exceptions to this agreement are the extraction of the non-hydrodynamic quadrupole damping rate $\Gamma_{Q,1}$, as well as the quadrupole frequency $w_Q$ for $\omega_\perp \tau_R\gtrsim 0.5$, which are qualitatively similar, but quantitatively different in the analytic and numerical results (see text for details).
\label{fig:one}}
\end{figure}

A different solution to Eq.~(\ref{eq:resBoltz}) is given by a scaling ansatz \cite{Dusling:2011dq},
\begin{equation}
\label{eq:anaini}
f(\bar t,{\bf x},{\bf \bar v})=f_{0}\left(\frac{\bar x_i}{b_i(\bar t)},\frac{\bar v_i-\bar x_i b_i^\prime(\bar t)/b_i(\bar t)}{\theta_i^{1/2}(\bar t)}\right)\,,
\quad 
b_i^{\prime\prime}(\bar t)+b_i(\bar t)-\frac{\theta_i(\bar t)}{b_i(\bar t)}=0\,,\quad
\theta_i^\prime(\bar t)+2 \frac{b_i^\prime(\bar t)}{b_i(\bar t)}\theta_i(\bar t)=-\frac{\theta_i-\bar\theta}{\tau_R \omega_\perp}\,,
\end{equation}
where $\bar \theta \equiv \frac{1}{D}\sum_{i=1}^D\theta_i$ (note that this solution is valid also for $D\neq 2$). In the limit of small perturbations $b_i(\bar t)=1+\delta b_i(\bar t)$, one can separate the equations into a breathing mode $\delta B(\bar t)=\frac{\delta b_x(\bar t)+\delta b_y(\bar t)}{2}$ and a quadrupole mode $\delta Q(\bar t)=\frac{\delta b_x(\bar t)-\delta b_y(\bar t)}{2}$. Using furthermore initial conditions with $\delta b_i'(0)=0$ one finds
\begin{equation}
\label{eq:lineq}
\delta B''(\bar t)+4 \delta B(\bar t)=0\,,\quad
\delta Q''(\bar t)+2 \delta Q(\bar t)+\tau_R \omega_\perp\left(\delta Q'''(\bar t)+4 \delta Q'(\bar t)\right)=0\,.
\end{equation}
From these linear equations, one clearly can identify an undamped ($\Gamma_B=0$) breathing mode oscillation 
\begin{equation}
\label{eq:anabreath}
\delta B(\bar t)=\beta e^{-\Gamma_B \bar t}\cos(w_B \bar t+{\rm const})
\end{equation}
with frequency $w_B=2$ (in units of the base frequency $\omega_\perp$) and a quadrupole mode with a frequency $w_Q$ that is $\sqrt{2}$ in the hydrodynamic limit $\tau_R \omega_\perp \rightarrow 0$, while it increases to the same frequency as the breathing mode in the free streaming limit $\tau_R \omega_\perp \rightarrow \infty$. A fully analytic solution to Eq.(\ref{eq:lineq}) for constant $\omega_\perp \tau_R$ is straightforward, and given by
\begin{equation}
\label{eq:anaquad}
\delta Q(\bar t)=\gamma e^{-\Gamma_{Q,0} \bar t}\cos(w_Q \bar t+{\rm const})+\delta e^{- \Gamma_{Q,1} \bar t}\,,
\end{equation}
where the explicit form for $\Gamma_{Q,0},\Gamma_{Q,1},w_Q$ is lengthy.
A plot of these quantities as a function of $\omega_\perp \tau_R$ is given in Fig.\ref{fig:one} and we note that the asymptotic behavior is
\begin{eqnarray}
\omega_\perp \tau_R &\ll& 1: \Gamma_{Q,0}\simeq \omega_\perp \tau_R\,,\quad
\Gamma_{Q,1}\simeq \frac{1-2 \omega_\perp^2 \tau_R^2}{\omega_\perp \tau_R}\,,\quad
w_Q\simeq \sqrt{2}+\frac{3 \omega_\perp^2 \tau_R^2}{2 \sqrt{2}}\,,\nonumber\\
\omega_\perp \tau_R &\gg& 1: \Gamma_{Q,0}\simeq \frac{1}{4\omega_\perp \tau_R}\,,\quad
\Gamma_{Q,1}\simeq \frac{1}{2 \omega_\perp \tau_R}\,,\quad
w_Q\simeq 2-\frac{5}{64 \omega_\perp^2 \tau_R^2}\,.
\label{eq:asym}
\end{eqnarray}

Inspecting Eq.~(\ref{eq:anaquad}), it becomes clear that it is a superposition of two different modes: a well-known 'hydrodynamic quadrupole mode' which is becoming dominant in the hydrodynamic limit $\omega_\perp \tau_R\rightarrow 0$ and a 'non-hydrodynamic' purely damped mode. The non-hydrodynamic mode is a feature that is common to evolution equations beyond Navier-Stokes (such as second-order hydrodynamics, cf.~\cite{Romatschke:2009im,Chao:2011cy}).

Because the derivation of Eq.(\ref{eq:resBoltz}) was carried out close to the hydrodynamic limit, we do not expect the analytically calculated value of $\Gamma_{Q,1}(\omega_\perp \tau_R)$ to be quantitatively reflected in any experimental measurement. Similarly, because the LB framework truncates the continuum Boltzmann equation onto a finite number of basis functions, we do not expect the numerical LB result to match the analytic value for $\Gamma_{Q,1}$, either. However, because the derivation is still qualitatively sound in this regime, we expect a term such as $A e^{-\Gamma_{Q,1} t}$ to also be present and observable in both the numerical LB simulation and experiments of trapped atomic clouds. Furthermore, we argue that extraction of the coefficient $\Gamma_{Q,1}(\omega_\perp \tau_R)$ from data could be very interesting because it is this coefficient which will indicate the radius of convergence of the hydrodynamic approximation. This is evident from recent progress in the context of relativistic fluid dynamics, where a term such as $A e^{-\Gamma_{Q,1} t}$ with non-hydrodynamic dependence on the relaxation time (\ref{eq:asym}) indicates the presence of so-called quasi-normal mode behavior \cite{Kovtun:2005ev,Heller:2013fn} (while there is only one such term in Eq.(\ref{eq:anaquad}), in practice we expect experimental signals to contain an infinite series of terms of the form $e^{-\Gamma_{Q,n} t}$, possibly also with oscillating components, with $\Gamma_{Q,n}\propto n$ for $n\gg 1$ originating from an infinite tower of quasinormal modes). Thus, by experimentally determining $\Gamma_{Q,1}$ (or possibly also information about higher order quasinormal modes) one can expect to learn about non-hydrodynamic behavior in strongly-coupled quantum fluids and we encourage experimentalists to consider this option in future work.

\subsection{2D Ideal Gas in an Harmonic Trap: Numerical Simulation}
\label{sec:idnum}

\begin{figure}[t]
\centering
\includegraphics[width=\textwidth]{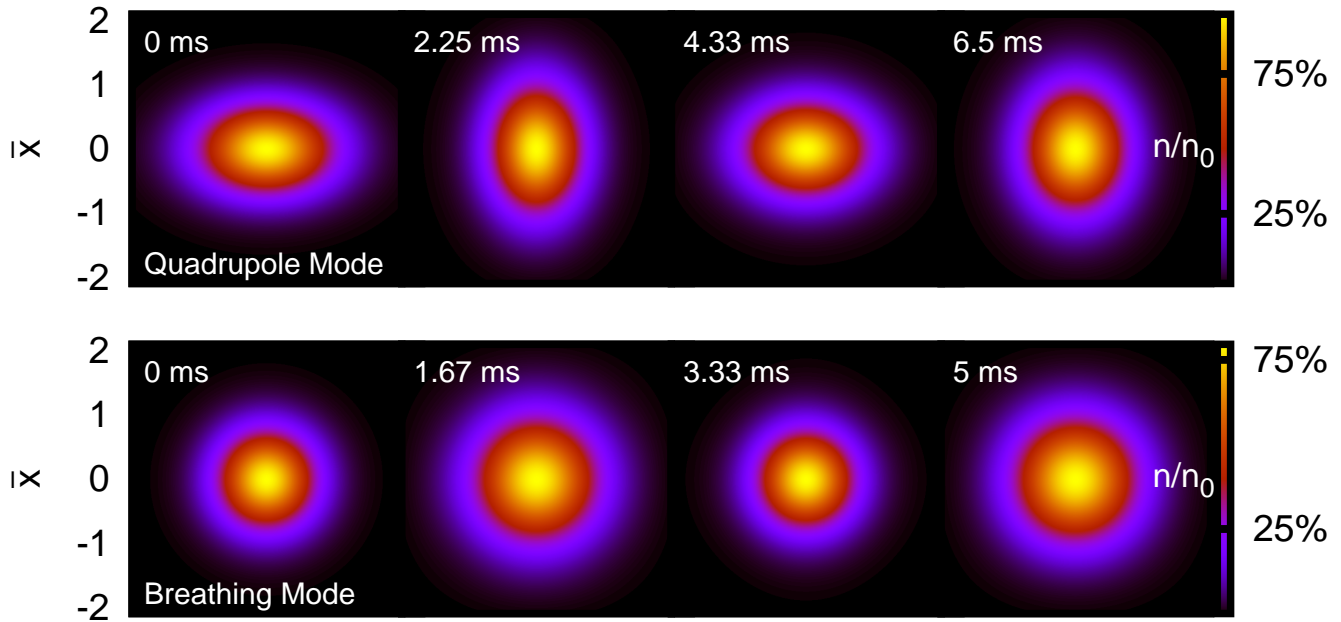}
\caption{Two-dimensional snapshots of the cloud density profile $n(\bar x,\bar y)$ (normalized by the central density $n_0$) for the breathing and quadrupole mode simulations. Time stamps are calculated assuming a transverse trapping frequency of $\omega_\perp=2\pi\times 125$ Hz \cite{Vogt:2011np}.
\label{fig:three}}
\end{figure}

In the previous subsection we have considered the analytic solution to the breathing and quadrupole modes of a two-dimensional Fermi gas in an harmonic trap. Here we proceed to simulate this setup using the Lattice Boltzmann algorithm outlined in section \ref{sec:lbone}. As advertised, we discretize the two-dimensional space on a square grid with lattice spacing $\delta \bar t$ such that $\bar x=i\, \delta \bar t$ with $i=1\ldots N$ where $N$ is the number of gridpoints we simulate along one dimension. Thus, the infinite volume limit and continuum limit of our simulation correspond to taking $N\rightarrow \infty$ with $\delta \bar t={\rm const}$ and $\delta\bar t\rightarrow 0$ with $N \delta\bar t={\rm const}$, respectively. In the results we present in the following, we have performed multiple simulations for different volumes and resolutions; while the infinite volume and continuum limit can never be reached in practice, we have striven to obtain robust extrapolation of our results to these limits, and we report the residual error from the extrapolation procedure in our LB results (see appendix \ref{sec:app} for details about the numerical procedure).

We first initialize the simulation in a configuration corresponding to the analytically tractable case presented in section \ref{sec:ana}, that is, a harmonic trapping potential with small oscillations around the equilibrium configuration. To this end, we define the amplitude $b_i(t)$ as the distance from the center where the density has dropped by a factor $e^{-1}$, and  we initialize the particle distribution function according to (\ref{eq:f0sol},\ref{eq:anaini}) with $\delta b_i(0)\simeq 10^{-2}$ or smaller. We then track the time evolution of the combinations $\delta B(\bar t),\delta Q(\bar t)$ and perform a fit of this evolution using the analytically derived model functions (\ref{eq:anabreath},\ref{eq:anaquad}), obtaining best-fit values for the coefficients $w_B,w_Q,\Gamma_B,\Gamma_{Q,0},\Gamma_{Q,1}$ as a function of $\omega_\perp \tau_R$ in the process. The extracted values for these coefficients are displayed in Fig.\ref{fig:one} along with the analytically calculated results. As can be seen from this comparison, the numerical simulation is in good quantitative agreement with the analytic results for the coefficients $w_B,\Gamma_B,\Gamma_{Q,0}$, even for values of $\omega_\perp\tau_R$ which are outside the hydrodynamic regime. The quadrupole mode frequency $w_Q$ agrees very well with the analytic result in the hydrodynamic limit, but starts to differ noticeably for values of $\omega_\perp \tau_R\gtrsim 0.5$. We attribute this disagreement to the discretization procedure of continuous velocities onto the D2Q25 grid, which implies that the solution to the LB equation only corresponds to the solution of the continuum Boltzmann equation in the hydrodynamic limit. We plan to test this hypothesis in future work by employing discretization grids with a larger number of velocity vectors, which should lead to $w_Q$ results that are closer to the analytic values.

As anticipated, the (non-hydrodynamic mode) damping rate $\Gamma_{Q,1}$ is found to be qualitatively similar to the analytic result, but in clear quantitative disagreement. Also, note that since $\Gamma_{Q,1}$ becomes very large in the hydrodynamic regime compared to the hydrodynamic damping $\Gamma_{Q,0}$, it becomes more and more difficult to extract $\Gamma_{Q,1}$ from the numerical simulation in the limit of $\omega_\perp \tau_R \lesssim 1$. We expect this to happen also in an experimental setup. However, by inspecting the analytic solutions given in Eqns.~(\ref{eq:anabreath}),(\ref{eq:anaquad}), one could try to engineer initial conditions for the cloud that would maximize the amplitude $\delta$, thus presumably leading to a better signal to noise ratio for extracting $\Gamma_{Q,1}$ in the hydrodynamic limit. We intend to pursue this direction in a follow-up study.

To summarize, our numerical LB simulation is able to accurately reproduce the collective behavior of a cold atomic gas cloud in the limit of small amplitude oscillations. This should be considered a successful test of the method. In the following, we will now use the numerical LB simulation to study the bulk evolution of a cold atomic gas cloud for situations where an analytic treatment is either not possible or difficult.

\subsection*{Interacting 2D Fermi Gas in a Gaussian Trap: Numerical Simulation}

In actual experimental setups, the trapping potential is usually not harmonic. In the case of 2D Fermi Gases, it is more accurately described by a Gaussian potential (cf.~\cite{2008PhRvA..78e3609R})
\begin{equation}
U({\bf \bar x})=V_0 \left(1-e^{-\frac{{\bf \bar x}^2}{\sigma^2}}\right)\,,
\end{equation}
where the potential depth is $V_0$ and the parameter $\sigma$ is related to the laser beam waist. If this potential is meant to approximate an harmonic trap close to the center ${\bf \bar x}\simeq 0$, then $\sigma^2=\frac{2c_L^2 V_0}{T_0}$. Fixing $\sigma$ in this way, the Gaussian trap corresponds to a one-parameter generalization of the harmonic trap, with $\frac{V_0}{T_0}$ controlling the degree of anharmonicity (the case of the purely harmonic trap is recovered in the limit $\frac{V_0}{T_0}\rightarrow \infty$). 

For an ideal gas, the equation of state takes the form $P(n,T)=n T$, and the results in sections \ref{sec:ana}, \ref{sec:idnum} have been obtained by using this (idealized) equation of state. For an ideal (meaning non-interacting) Fermi gas, the equation of state is different from that of an ideal gas because the Fermi statistics imply a non-linear relation between the pressure and the density. Furthermore, in setups relevant for cold atom experiments at unitarity, the equation of state is known to be different from both the ideal gas and ideal Fermi gas \cite{2012Sci...335..563K}, particularly for the two dimensional case $D=2$ \cite{PhysRevLett.112.135302}. In order to have a realistic description of the dynamics, we thus implement the interacting equation of state from Ref.~\cite{PhysRevLett.112.135302} in our simulations. The equation of state is constructed out of tabulated data for the density as a function of chemical potential $\mu$ and temperature $T$, $n=n(\mu/T)$. Results are available for various values of the physical binding energy of the two-body bound state $E_b$, which is always present for an attractive 2D Fermi gas \cite{LLQMbook,PhysRevLett.62.981}. From the density, the pressure can be calculated numerically through direct integration of the thermodynamic relation $n\equiv \left.\frac{\partial P(\mu,T)}{\partial \mu}\right|_{T}$, matched to the virial expansion for small densities:
\begin{equation}
P\simeq n T (1+B_2 n)\,,\quad \frac{\mu}{T}<-5\,.
\end{equation}

With the equation of state fixed, the initial conditions for an isothermal atomic cloud are corresponding to solutions to the equations of hydrostatics 
(cf. Eq.~(\ref{eq:rho})),
\begin{equation}
\nabla P(\mu,T)
=n \nabla \mu({\bf x})=-n \nabla U({\bf x})\,.
\end{equation}
Thus, a solution to the hydrostatic equations, in terms of rescaled coordinates
, is given by
\begin{equation}
\mu(\bar {\bf x})=\mu_0-U(\bar {\bf x})\,,
\end{equation}
where $\mu_0$, the chemical potential at the trap center, is sometimes referred to as the Fermi Energy. For fixed $\mu_0$ and temperature $T_0$, the total particle number $N$ is given by an integral over the number density,
\begin{equation}
\label{eq:Nvsn}
N=\int d^2{\bf x}\, n\left(\mu({\bf x},T)\right)\,.
\end{equation}
In the idealized case of harmonic trapping potential and non-interacting Fermi gas, Eq.~(\ref{eq:Nvsn}) simplifies in the zero-temperature limit as $N\rightarrow \frac{\mu_0^2}{2 \omega_\perp^2}$, 
which is often used to define an idealized ``Fermi Temperature'' $T_F=\mu_0$ of a trapped atomic gas as
\begin{equation}
\label{eq:TF}
T_F\equiv \sqrt{2 N} \omega_\perp\,.
\end{equation}

\begin{figure}[t]
\centering
\includegraphics[width=0.45\textwidth]{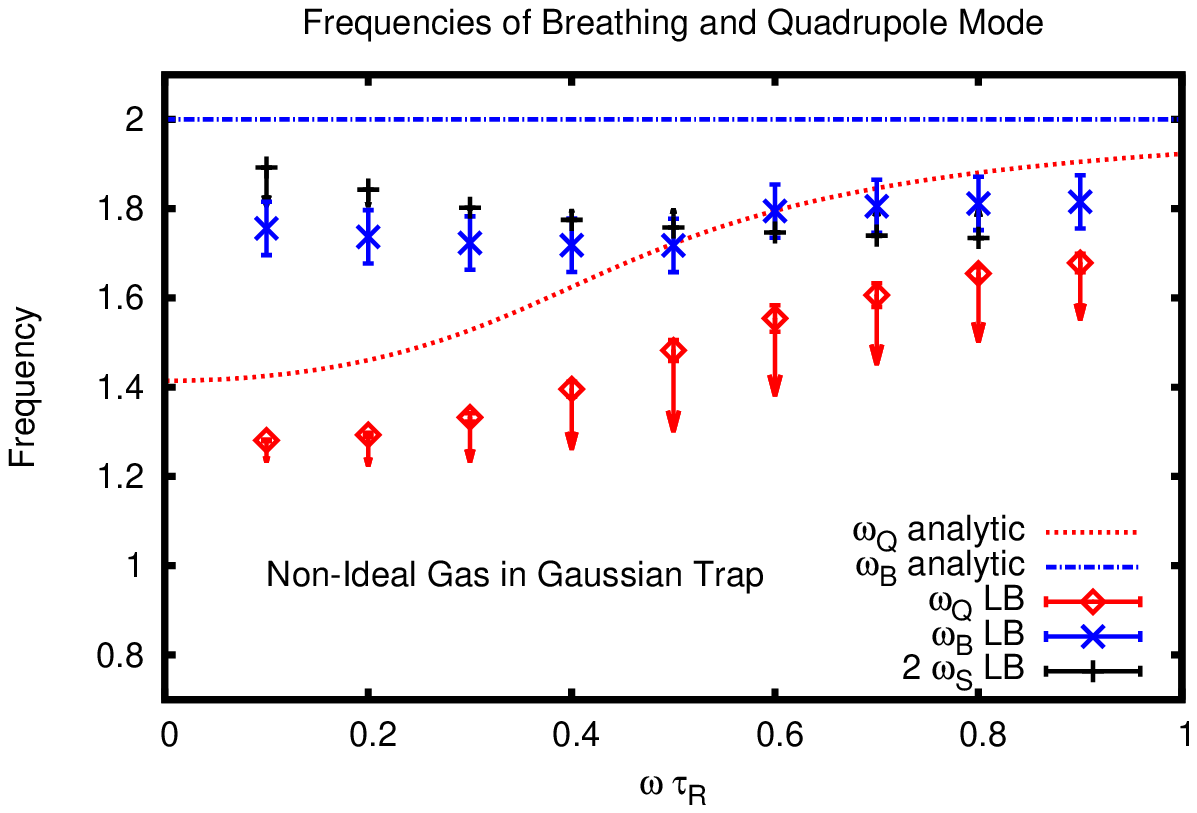}\hfill
\includegraphics[width=0.45\textwidth]{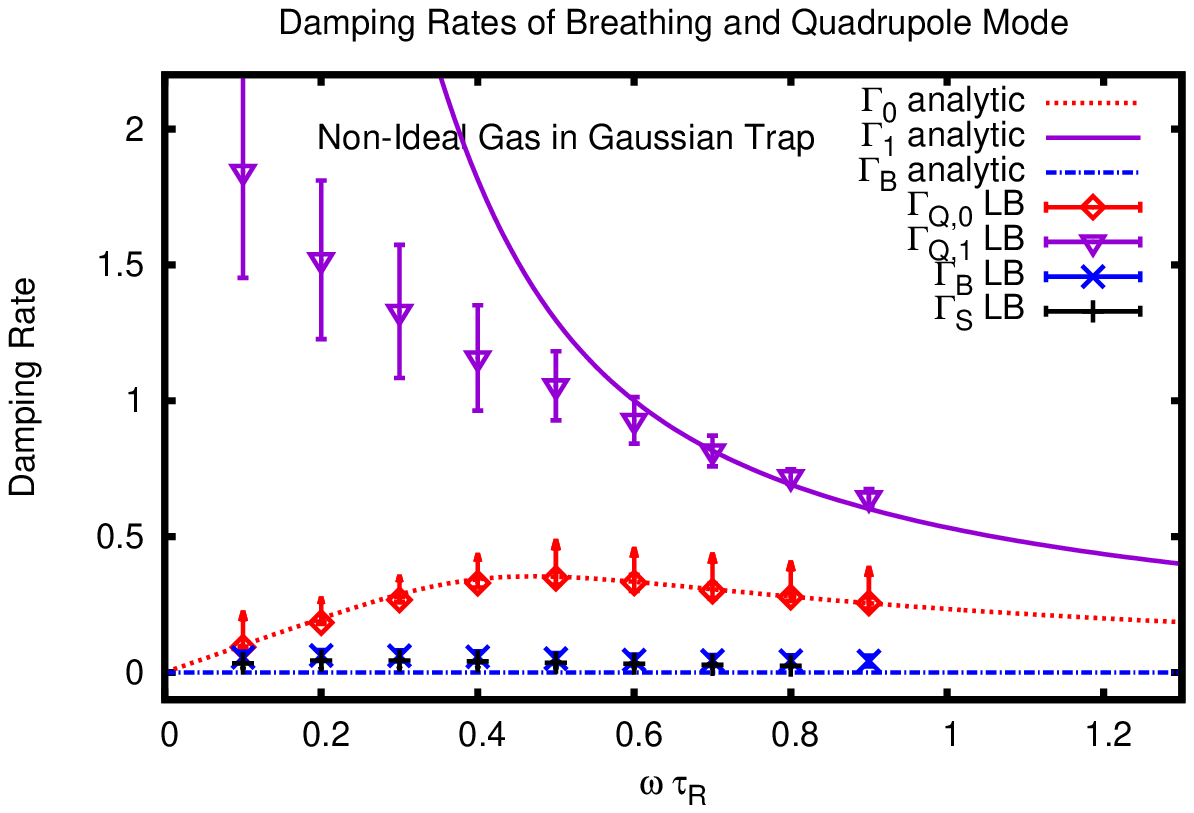}
\caption{Frequency and damping rates for the sloshing, breathing and quadrupole mode $c,\delta B,\delta Q$, respectively, for an interacting 2D Fermi gas in a Gaussian trap with $\frac{V_0}{T_0}=10$. Shown are fitted values from a fully numerical LB simulation (symbols with systematic errors from the infinite volume extrapolation where available; otherwise infinite volume trend is indicated by an arrow, see appendix \ref{sec:app} for details). For comparison, the analytic results for an ideal gas in a harmonic trap (cf. Fig.~\ref{fig:one}). We find that the extracted frequencies in the interacting, Gaussian trap case show the same qualitative behavior as the idealized analytic result as a function of $\omega_\perp \tau_R$, but are systematically lower. Furthermore, the extracted breathing mode damping rates $\Gamma_S,\Gamma_B$ in the interacting, Gaussian trap case are non-vanishing, in contrast to the idealized analytic result.  
\label{fig:GT}}
\end{figure}

In experiments on a cloud of trapped $^{40}K$ atoms close to unitarity, the temperature of the system is reported in units of $T_F$ as defined in Eq.~(\ref{eq:TF}), finding $\frac{T}{T_F}=0.37-0.9$ for an average number of atoms of $N\simeq 2000$ per 2D cloud \cite{Vogt:2011np}. Once the temperature and the number of atoms is known, the chemical potential $\mu_0$ at the center of an arbitrary trapping potential can be obtained by numerically inverting Eq.~(\ref{eq:Nvsn}), which is the strategy we employ in the following. To be explicit, we choose to simulate a 2D cloud of $N=2320$ atoms and $\frac{T_0}{T_F}=0.45$ in the following. At any instant in time, we assume the cloud to be isothermal, but we allow the temperature to fluctuate as a function of time $T=T(\bar t)$. Also, we choose an interaction strength corresponding to $\frac{E_b}{T_0}=1.0$, which can be related to the experimentally reported quantity $\ln(k_F a)$ as follows \cite{Tilmanpc}:
\begin{equation}
\label{eq:lnkfa}
\ln{(k_F a)}=-\frac{1}{2}\ln\left[\frac{T_0}{2 T_F} \frac{E_b}{T_0}\right]\simeq 0.74\,.
\end{equation}
A fully realistic implementation of the trapping potential would require precise knowledge of the laser beam waist parameters, gravitational effects as well as the magnetic field gradients (see e.g. the discussion in Ref.~\cite{2013PhRvA..88e3616C}). While we aim to implement this in a follow-up study, for the present work we chose a reference value $\frac{V_0}{T_0}=10$ that is comparable to experimental values in Ref.~\cite{Vogt:2011np}. 

The above choices of parameters imply a central chemical potential over temperature ratio of $\frac{\mu_0}{T_0}\simeq 0.1$. Note that for a non-ideal equation of state, the simulated viscosity over density ratio becomes (cf. Eq.~(\ref{eq:tauReta}))
\begin{equation}
\frac{\eta}{n}=\frac{P}{n T_F} \sqrt{2 N} \omega_\perp \tau_R\,,
\end{equation}
which is in general density and temperature dependent. For further reference, we note that $\frac{\eta}{n}\simeq 2$ for $\omega_\perp \tau_R=0.1$ at $\mu_0=0$. However, it should be emphasized that strictly speaking the viscosity of a two-dimensional fluid is ill-defined because of the presence of thermal fluctuations (cf. \cite{Chafin:2012eq,Romatschke:2012sf}), a topic which we intend to revisit in future work.

Once the physical parameters have been specified, we simulate the collective behavior of the cold atomic cloud using the LB algorithm outlined in Sec.\ref{sec:idnum}. We start with the initial condition of a slightly perturbed cloud in a Gaussian trap and track the time evolution of the sloshing, breathing and quadrupole modes, $c(\bar t),\delta B(\bar t),\delta Q(\bar t)$. 

Unlike in the case of ideal equation of state and harmonic trapping potential, we find signs of transient phenomena not captured by the solution structure given in Eqns.(\ref{eq:slosh},\ref{eq:anabreath},\ref{eq:anaquad}) in LB simulations at finite volume and resolution. These transient phenomena could be due to 'overtones' to the hydrodynamic sloshing, breathing and quadrupole modes, with higher frequencies and damping rates than the fundamental modes. However, our current numerical accuracy is not sufficient to distinguish real overtones from possible numerical artifacts at finite volume and resolution, so we intend to revisit this issue in a follow-up high precision study.

In general, one finds that the presence of an anharmonic trapping potential will affect the evolution equations for the sloshing, breathing and quadrupole mode, resulting in a change of the frequencies $w_S,w_Q,w_B$ with respect to the harmonic trap values Eq.~(\ref{eq:asym}). This effect has been noted before \cite{2004PhRvA..70e1401K}.

The extracted frequencies $w_S,w_B,w_{Q}$ and damping rates $\Gamma_S,\Gamma_{B},\Gamma_{Q,0},\Gamma_{Q,1}$ from the LB simulation of collective modes of a two-dimensional non-ideal Fermi gas in a Gaussian trap are shown in Fig.~\ref{fig:GT}. The behavior of the frequencies and damping rates are qualitatively similar to the case of the idealized, harmonic trap case, but shifted to lower values. Since in the experimental setup the sloshing mode frequency is used to calibrate $\omega_\perp$, such a frequency shift is not apparent in the experimental measurements \cite{Vogt:2011np}.
Moreover, one finds that in the Gaussian potential trap both the breathing mode and sloshing mode damping rates $\Gamma_B,\Gamma_S$ are no longer consistent with zero, but found to be a small but non-vanishing value, similar to what has been found in experiment \cite{Vogt:2011np}. 

\begin{figure}[t]
\centering
\includegraphics[width=0.45\textwidth]{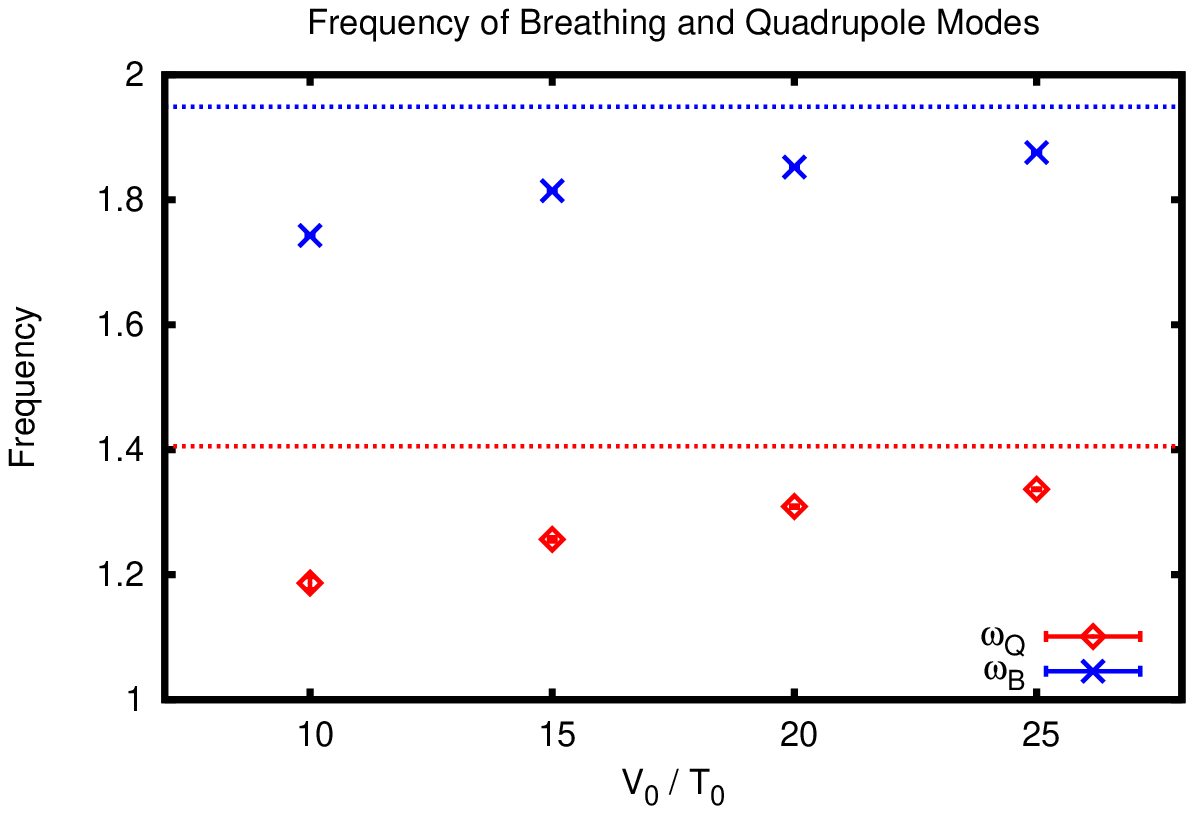}\hfill
\includegraphics[width=0.45\textwidth]{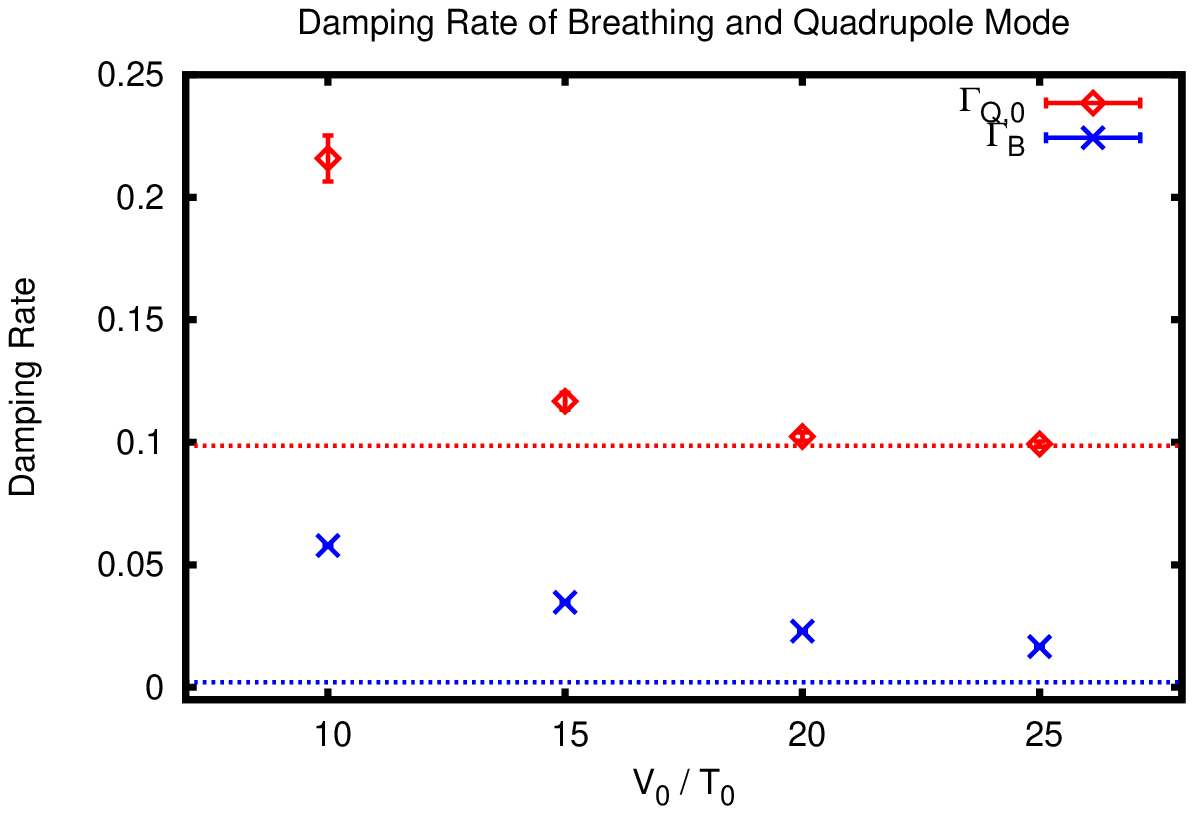}
\caption{Frequency and damping rates for the breathing and quadrupole mode $\delta B,\delta Q$, respectively, for an interacting 2D Fermi gas with $\omega_\perp \tau_R=0.1$ in a Gaussian trap varying anharmonicity parameter $\frac{V_0}{T_0}$. Shown are fitted values from a fully numerical LB simulation (symbols with systematic errors from the infinite volume extrapolation). Dotted lines indicate the results for $\frac{V_0}{T_0}=100$. In the harmonic trap limit (corresponding to $\lim \frac{V_0}{T_0}\rightarrow \infty$), one essentially recovers the idealized results shown in Fig.~\ref{fig:one}, indicating that the non-ideal equation of state does not strongly affect either quadrupole or breathing mode in the LB simulations. Not shown is the non-hydrodynamic quadrupole mode damping $\Gamma_{Q,1}$, because for $\omega_\perp \tau_R=0.1$ this mode is strongly damped and could not be unambiguously extracted from the simulations.
\label{fig:trapdep}}
\end{figure}

\begin{figure}[t]
\centering
\includegraphics[width=0.45\textwidth]{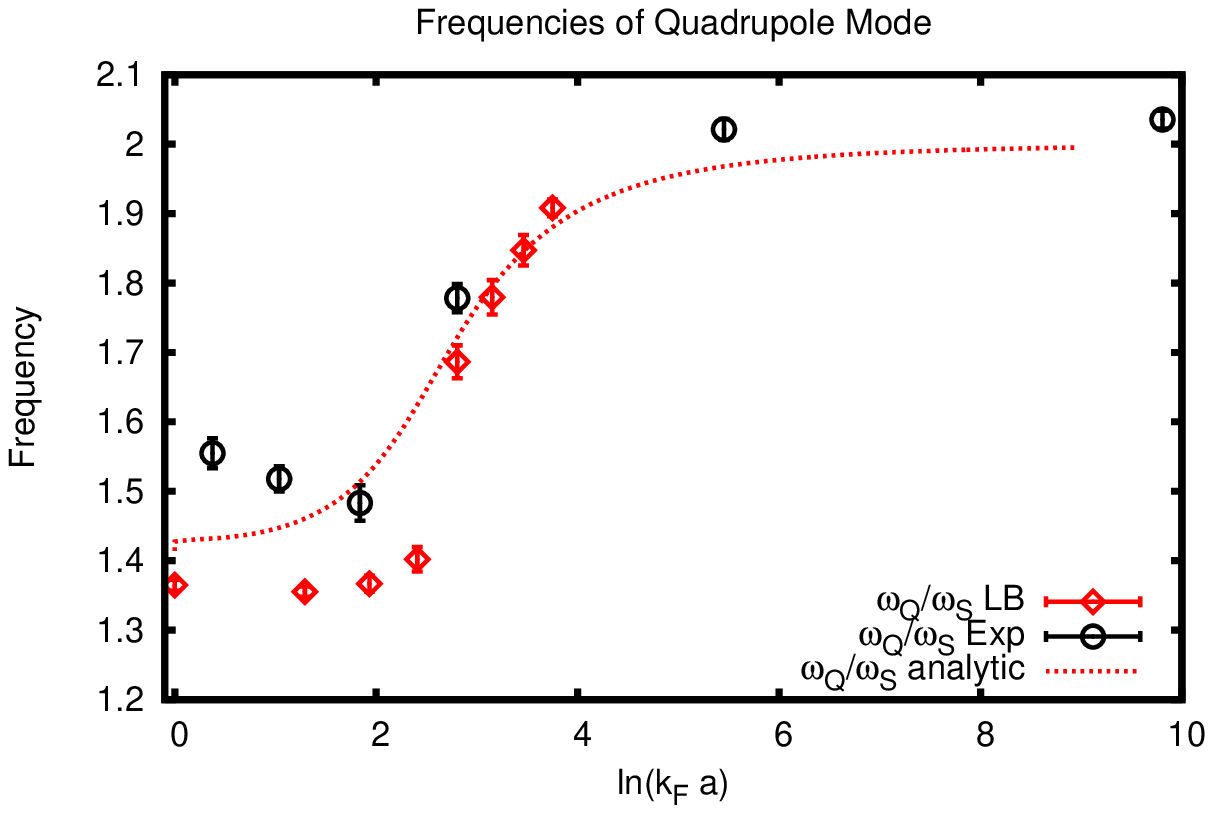}\hfill
\includegraphics[width=0.45\textwidth]{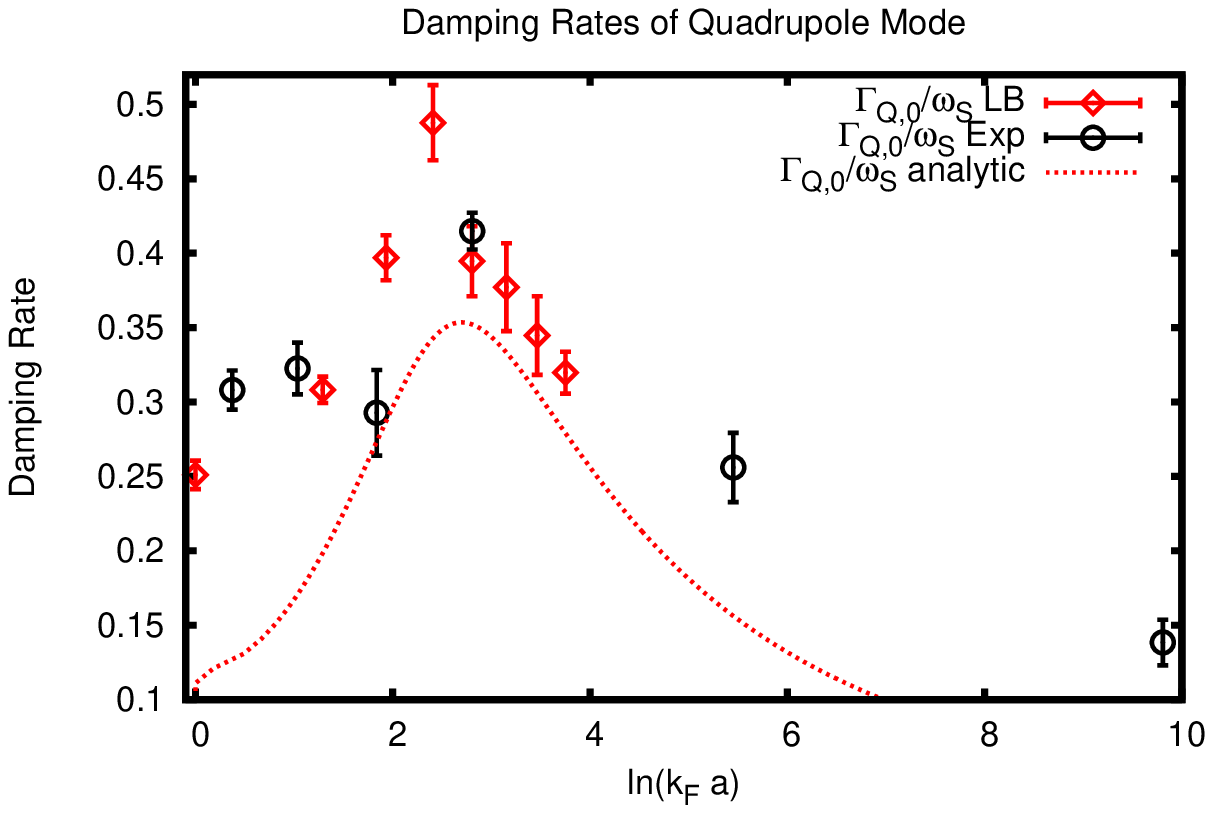}
\caption{Quadrupole frequency and damping rate (normalized by the sloshing mode frequency) measured in Ref.~\cite{Vogt:2011np} ('Exp') compared to the results of the LB simulation with anharmonicity parameter $\frac{V_0}{T_0}=10$ ('LB') and the analytic result using the one-parameter fit given in Eq.~(\ref{eq:fit}).
\label{fig:five}}
\end{figure}

Indeed, when considering the harmonic trap limit $\lim \frac{V_0}{T_0}\rightarrow \infty$, the results for all modes, sloshing, breathing and quadrupole mode, tend to the result for an ideal gas in an harmonic trap shown in Fig.~\ref{fig:one}. This suggests that any non-ideal equation of state effects have a minor impact on the extracted frequencies and damping rates of the modes considered here. Thus, the frequencies and damping rates are essentially controlled by the value of $\omega_\perp \tau_R$ as well as the trap anharmonicity parameter $\frac{V_0}{T_0}$. In view of this we can attempt to compare the extracted frequencies and damping rates in our LB simulations to experimentally determined values, as has been done before by other authors (cf.~\cite{Schafer:2011my,2012PhRvA..85a3636B,PhysRevA.85.045601,PhysRevA.86.013617}). To do this, we need to match the value of $\omega_\perp \tau_R$ to the interaction strength. At finite interaction strength, it is reasonable to assume that the relaxation time $\tau_R$ is proportional to the inverse of the imaginary part of the scattering amplitude (cf.~\cite{Vogt:2011np,2012PhRvA..85a3636B}), thus
\begin{equation}
\label{eq:fit}
\omega_\perp \tau_R = K \left(1+\frac{4}{\pi^2}\ln^2(k_F a)\right)\,,
\end{equation}
with $K$ a (density and temperature dependent) normalization factor. An average value of $K$ can be estimated by matching the maximum of the quadrupole damping rate found in the LB simulations at $\omega_\perp \tau_R \simeq 0.5$ to the location at $\ln (k_F a)\simeq 3$ found in the experiment of Ref.~\cite{Vogt:2011np}, giving $K\simeq 0.12$. The comparison between LB simulation with $\frac{V_0}{T_0}=10$ of the quadrupole damping rate and frequency using Eq.~(\ref{eq:fit}) is shown in Fig. \ref{fig:five}. From this comparison, it can be seen that with the one-parameter fitting through Eq.~(\ref{eq:fit}), the overall agreement between the LB simulation and experimental values for the frequency and damping rate can be considered reasonable. However, note that the larger LB damping rate in Fig. \ref{fig:five} compared to the analytic result stems mainly from the strong anharmonicity effects encountered for $\frac{V_0}{T_0}=10$ (cf. Fig.~\ref{fig:trapdep}), which may be larger than in the actual experimental setup. On the other hand, not included in Fig. \ref{fig:five} are effects of density-dependent $\omega \tau_R$, which are expected to increase the damping rate with respect to the analytic result \cite{PhysRevA.86.013617}. We intend to perform a more detailed comparison to experimental data, including an investigation of the above points as well as the effects of changing the temperature, in a follow-up study.

\section{Conclusions}
\label{sec:conc}

In this work, we have presented fully nonlinear dissipative fluid dynamics simulations of a trapped two-dimensional Fermi gas close to unitarity based on the Lattice Boltzmann algorithm. We were able to verify our simulations using the analytically tractable case of an ideal gas in an harmonic trap, finding excellent agreement in the fluid dynamics regime, as well as qualitative agreement in the ballistic (non-interacting) regime. Furthermore, we were able to simulate the case of non-ideal equations of state as well as non-harmonic trapping potentials, relevant to the study of collective modes in cold atom experiments. For convenience, we have made our simulation source code publicly available at \cite{codedown}.

Based on our simulations as well as analytic results, we predicted the presence of a non-hydrodynamic component of the quadrupole collective mode, which should be observable in experiments.

We expect our work to be a step towards a fully realistic simulation of trapped Fermi and Bose gases in two and three spatial dimensions, close to (but not limited to) the unitary regime. Our study is complementary to other realistic simulation approaches, such as those of Refs.~\cite{Pantel:2014jfa,Bluhm:2015raa,2013PhRvA..88e3616C,2014JPhCS.497a2028U,0295-5075-97-1-16003,PhysRevA.85.045601}. We believe that comparing and combining these simulation results will open up the possibility of precision determination of transport coefficients from experiments of cold atomic gases, such as the shear and bulk viscosities and heat conductivity. 

\begin{acknowledgments}
 
This work was supported in part by the Department of Energy, DOE award No. DE-SC0008132. We are indebted to Tilman Enss for providing us with tabulated data for the equation of state of a two-dimensional Fermi gas at different interaction strengths from Ref.~\cite{PhysRevLett.112.135302} and to Marco Koschorreck for sending us raw experimental data for the breathing mode evolution from Ref.~\cite{Vogt:2011np}. Furthermore, we would like to thank Tilman Enss and Ana Maria Rey for many fruitful discussions on this topic.

\end{acknowledgments}

\begin{appendix}
\section{Details on Numerical Scheme: Infinite Volume, Continuum Limit, Conserved Quantities}
\label{sec:app}

\begin{figure}[t]
\centering
\includegraphics[width=0.7\textwidth]{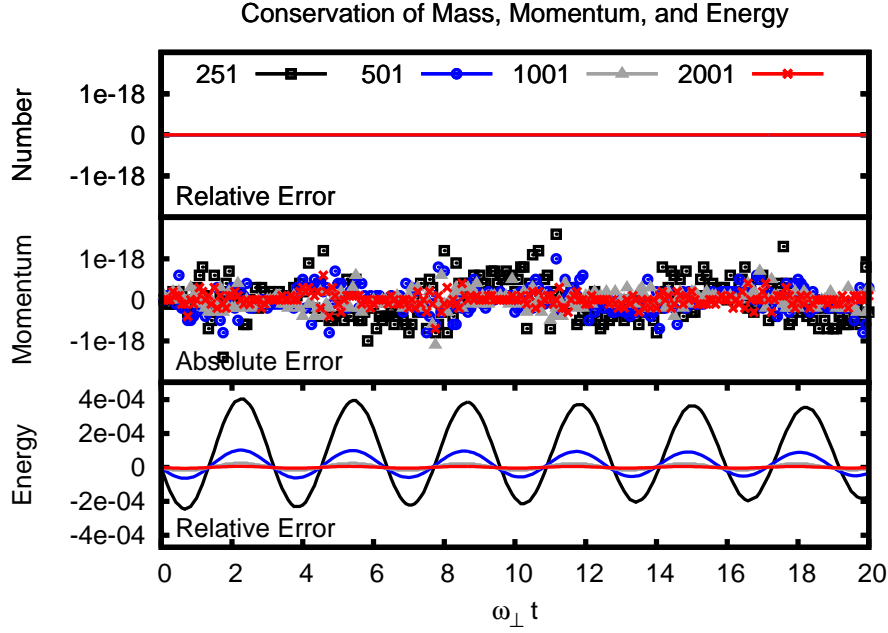}\hfill
\caption{\label{fig:ap1}
Time evolution of total particle number $N(t)$, total momentum $P_x(t)$ ($P_y(t)$ is similar) and total energy $E(t)$. Shown is relative error $N(t)/N(0)-1$, $E(t)/E(0)-1$ for particle number and energy and absolute error $P_x(t)$ for total momentum for various resolutions (parametrized by the number of gridpoints $N=251,501,1001,2001$). 
}
\end{figure}

In this appendix, we give details about the precision of our numerical scheme, as well as the infinite volume and continuum limit. All simulation results shown here are for harmonic trapping potential and $\tau_R\omega_\perp=0.1$.

In Fig.~\ref{fig:ap1}, we consider the time-evolution of the total particle number, total momentum and total energy in our simulation. All these quantities should be exactly conserved, but as is generally the case with numerical schemes, conservation is broken in the numerical evolution. This is not a problem as long as the violation of conserved quantities is small in overall magnitude and converging to zero in the infinite volume and continuum limit. Fig.~\ref{fig:ap1} highlights the relative error for the total number density and energy conservation, and the absolute error for the momentum conservation for a fixed volume $L=N \delta \bar t=\frac{251}{60}$ and increasingly better resolution (parametrized by increasing $N$ from 251 to 2001 points). As can be seen in Fig.~\ref{fig:ap1}, our scheme conserves particle number and total momentum to machine precision, while energy conservation is dominated by resolution artifact. Note that in order to achieve exact energy conservation, one would need to include polynomials up to fourth order in Eq.~(\ref{eq:BE}). In the case at hand, the energy conservation violation is not increasing in magnitude as a function of simulation time, so our simulations are long-time stable. Also, energy violations
converge to zero quadratically with resolution ${\cal O}(\delta \bar t^2)$, so that our scheme is second-order convergent to the exact energy conservation limit.

\begin{figure}[t]
\centering
\includegraphics[width=0.7\textwidth]{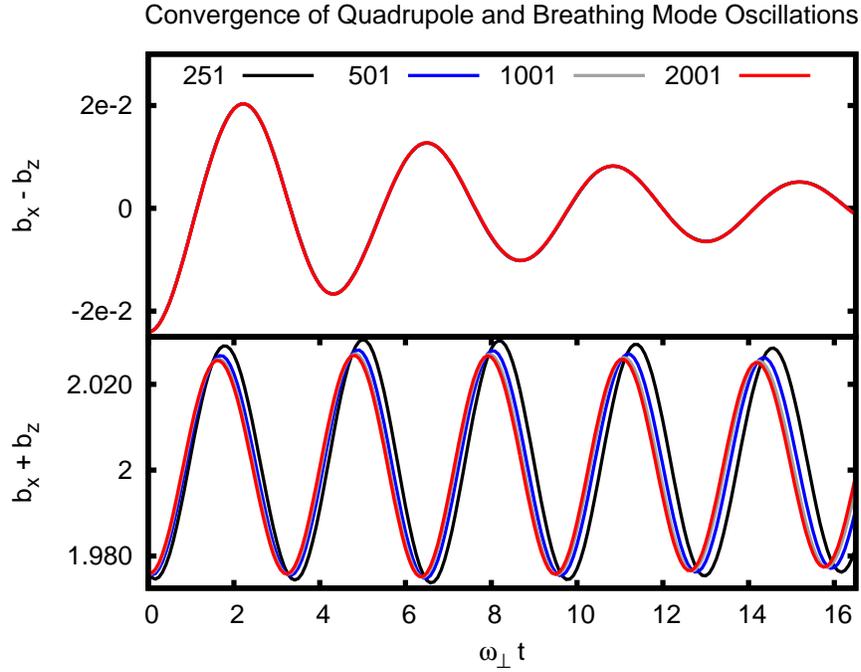}\hfill
\caption{\label{fig:ap2}
Continuum limit convergence of the time evolution for quadrupole mode ($b_x-b_z$) and breathing mode ($b_x+b_z$) for fixed volume and increasing resolution (parametrized by the number of gridpoints $N$ along one dimension $N=251,501,1001,2001$).}
\end{figure}

In Fig.~\ref{fig:ap2}, the continuum limit $\delta \bar t\rightarrow 0$ with $N \delta \bar t={\rm const}$ of the quadrupole and breathing mode oscillations are studied. One finds that the quadrupole mode is fairly insensitive to finite resolution artifacts, whereas the breathing mode converges in second-order to the continuum limit. Note that even though the total amplitude of the simulated breathing mode is fairly sensitive to the resolution (the reason being the numerical correction terms in Eq.~(\ref{eq:numcorr})), neither the breathing mode frequency nor the damping rate show a strong sensitivity. This implies that reliable extractions of frequency and damping rate can be performed from simulations with rather coarse resolutions.

\begin{figure}[t]
\centering
\includegraphics[width=0.45\textwidth]{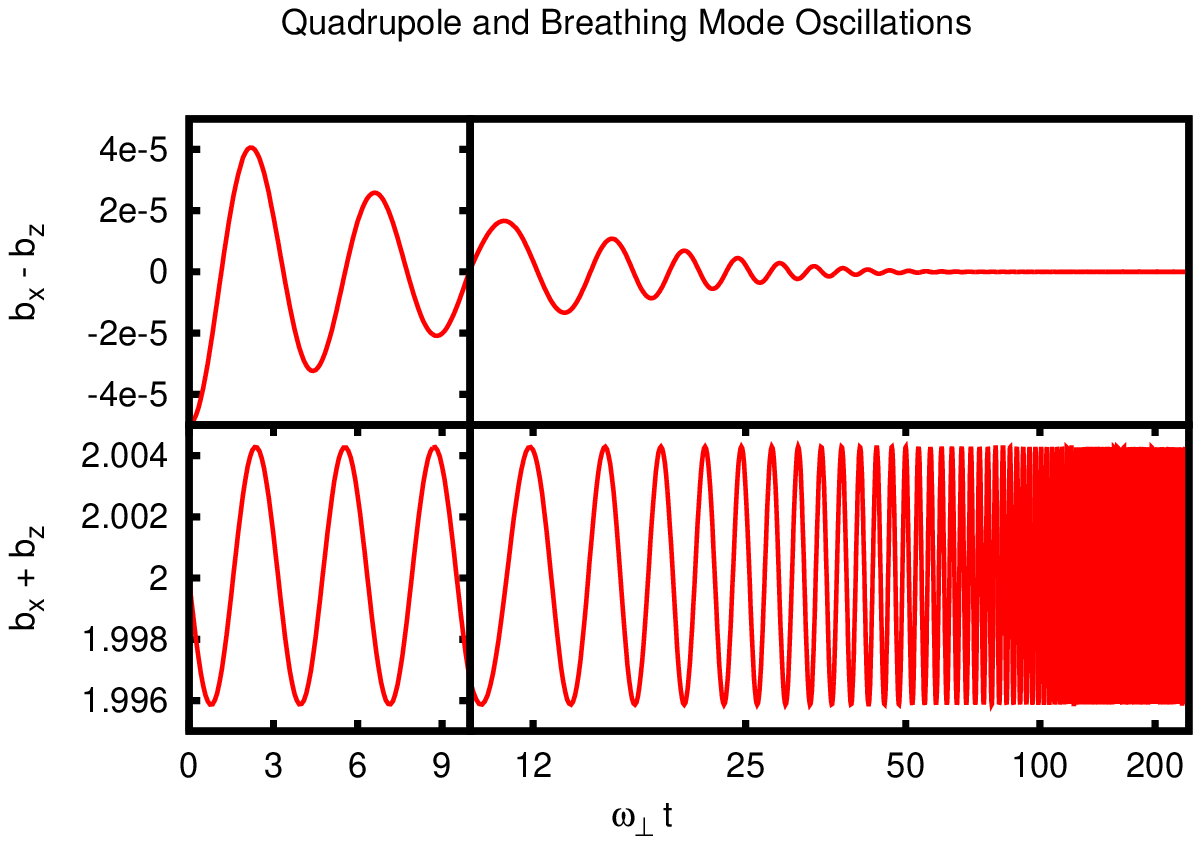}\hfill
\includegraphics[width=0.45\textwidth]{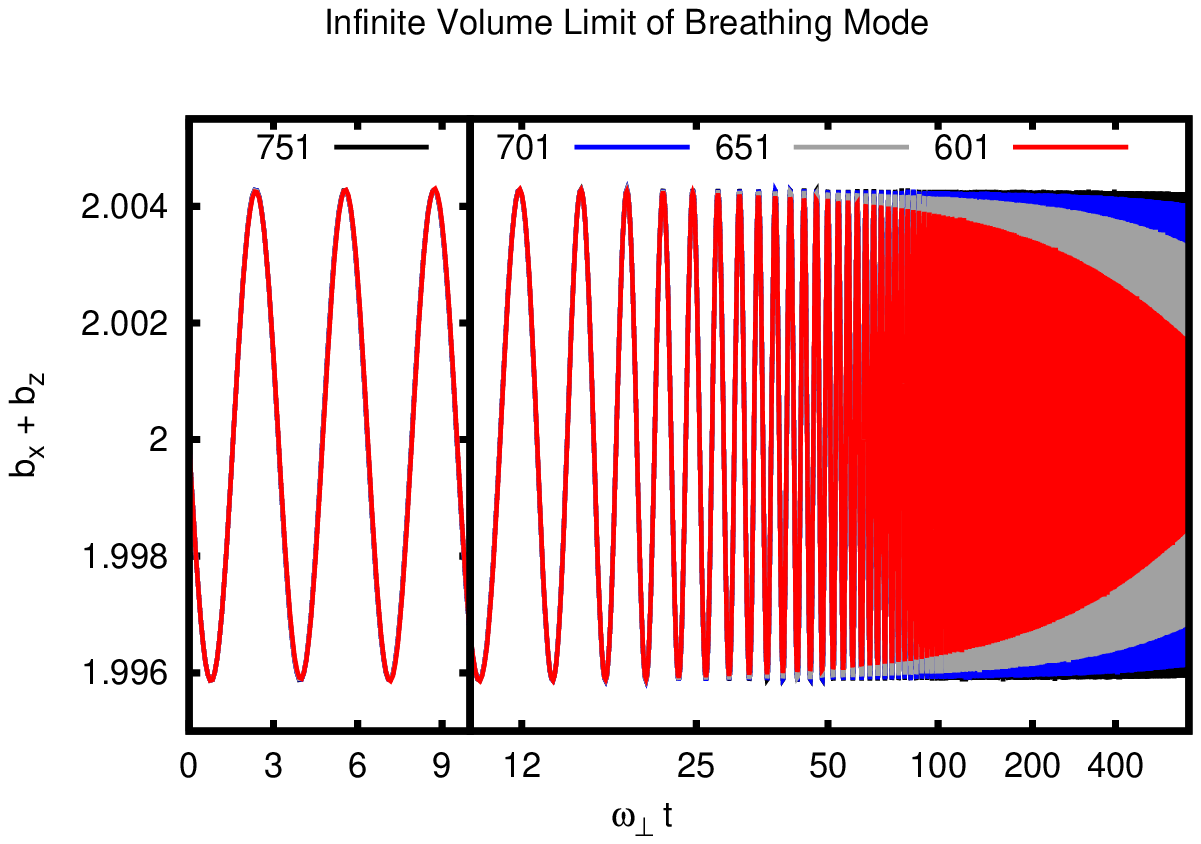}
\caption{\label{fig:ap3}
Left: Long-term evolution of quadrupole mode ($b_x-b_z$) and breathing mode ($b_x+b_z$) for $N=751,\delta \bar t=\frac{1}{120}$. Right: Infinite-volume limit for even longer time evolutions for the breathing mode for $\delta \bar t=\frac{1}{120}$ and $N=601,651,701,751$.}
\end{figure}

In Fig.~\ref{fig:ap3} we consider the long-term time evolution of the breathing and quadrupole mode for fixed resolution and various simulation volumes. As a consequence of conserving number, momentum and energy in our numerical scheme, our simulations remain stable essentially forever (note that for $\omega_\perp=2\pi\times 125$ Hz, $\omega_\perp t=300$ corresponds to $380$ ms, much longer than typically studied in experimental setups). Interestingly, as can be seen in the rhs panel of Fig.~\ref{fig:ap3}, we find that at very late times the breathing mode amplitude starts to decay. This decay is a finite volume artifact, and can be removed by simulating larger volumes (see again the rhs panel of Fig.~\ref{fig:ap3}), and we find that in the infinite volume limit ($\delta \bar t={\rm const}, N\rightarrow \infty$) the amplitude of the breathing mode is constant in an harmonic trapping potential, as expected from analytic results in continuum.

\begin{figure}[t]
\centering
\includegraphics[width=0.45\textwidth]{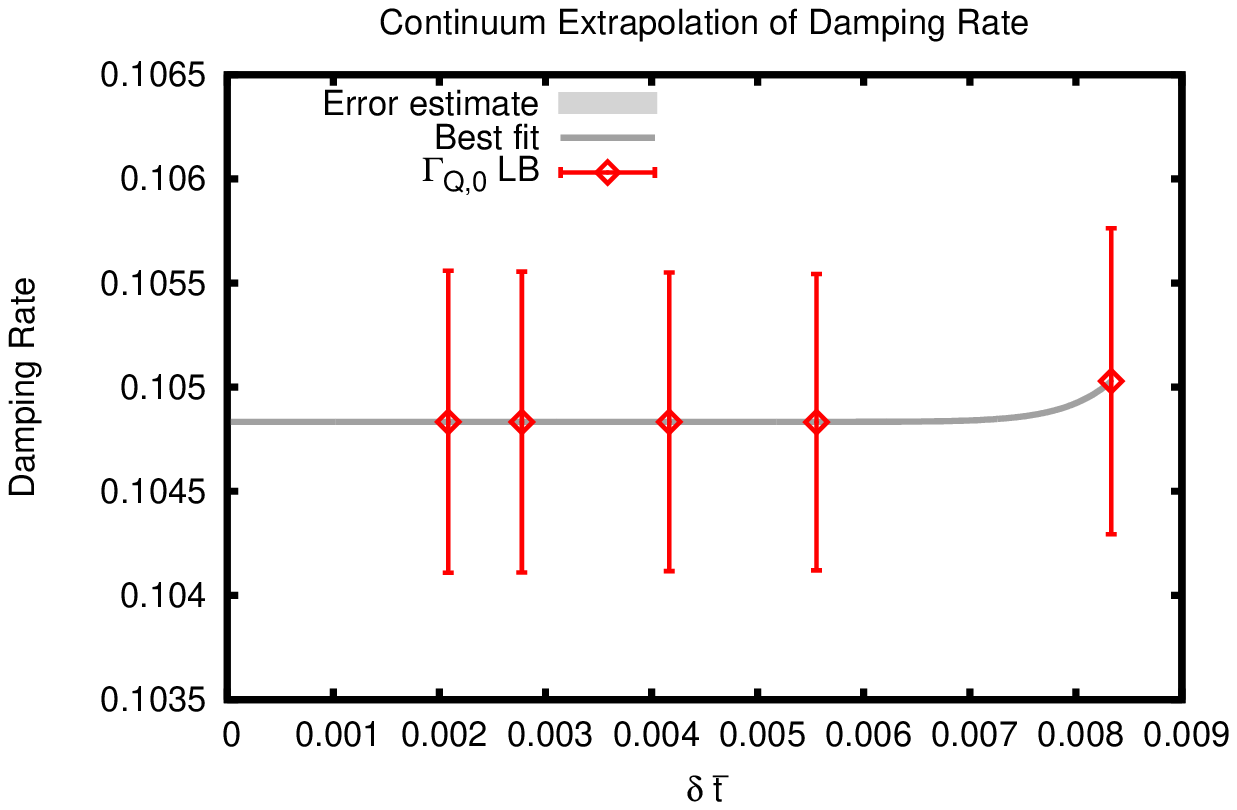}\hfill
\includegraphics[width=0.45\textwidth]{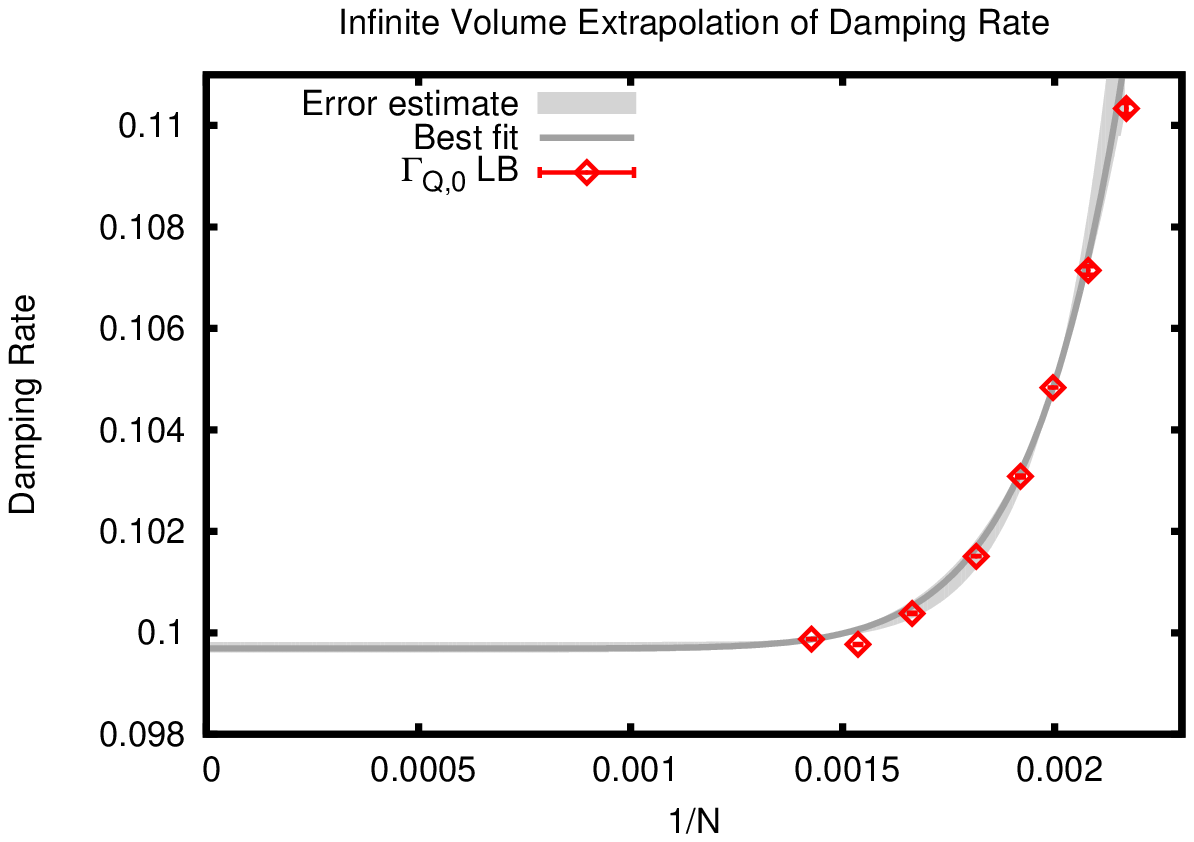}
\caption{\label{fig:ap4}
Left: Continuum extrapolation of the quadrupole mode damping rate $\Gamma_{Q,0}$ for constant volume $N \delta \bar t=\frac{501}{120}$. Right: infinite volume extrapolation of the continuum extrapolated damping rate values (error bars from continuum extrapolation are smaller than symbol sizes).}
\end{figure}

For results on the frequency and damping rate shown in the main text of this article we used multiple simulations at different volume and different resolution. Using these results we perform extrapolations to the infinite volume and continuum limit. To be more specific, we first performed simulations at fixed volume $N \delta \bar t={\rm const}$ and various resolutions $\delta \bar t$, and extracted damping rates and frequencies of interest for all of these simulations. To extrapolate to the continuum for we then performed one-parameter power-law least-square fits to the finite-resolution data. For example, using results for the quadrupole damping rate $\Gamma_{Q,0}(\delta \bar t)$ at various resolutions $\delta \bar t$ we obtain least-square fits of the form
\begin{equation}
\label{eq:cont-ex}
\Gamma_{Q,0}(N,\delta \bar t)=c_0(N)+c_1(N) (\delta \bar t)^{n}\,,\quad n=1,2,3,4,\ldots\,,
\end{equation}
and select those values of $n$ which have the overall smallest least-square value (``best fit'') and two more which have the second and third smallest least-square value (to quantify the quality of the fit and uncertainty of the extrapolation). The continuum extrapolated value for $\Gamma_{Q,0}$ is then found by evaluating the fit function (\ref{eq:cont-ex}) for the best-fit value of $n$ at $\delta \bar t=0$, and the uncertainty of the extrapolation is obtained by similarly evaluating the second and third-best fit function. The result of the procedure is shown for the case of $\tau_R \omega_\perp=0.1$ in the lhs of Fig.~\ref{fig:ap4}.

Once the continuum extrapolation has been performed for several volumes, we perform a min-$\chi^2$ fit to the continuum extrapolated data including error bars. For example, we use 
$$
\Gamma_{Q,0}(N^{-1},\delta \bar t=0)=d_0+d_1 N^{-n}\,, \quad n=1,2,3,4,\ldots\,,
$$
and select those values of $n$ which have the overall smallest $\chi^2$ value (``best fit'') and two more which have the second and third smallest $\chi^2$ value (to quantify the quality of the fit and uncertainty of the extrapolation to infinite volume). An example of the infinite volume extrapolation of the continuum extrapolated data is shown in the rhs of Fig.~\ref{fig:ap4}. 
After this extrapolation procedure we thus obtain the continuum and infinite volume extrapolated quantities such as $\lim_{\delta \bar t\rightarrow 0} \lim_{N\rightarrow \infty} \Gamma_{Q,0}(N,\delta \bar t)$, including uncertainty estimates from the extrapolation procedure. Wherever possible, we report results for these extrapolated quantities (rather than results at finite volume or finite resolution) in the main text of this article. 

\begin{figure}[t]
\centering
\includegraphics[width=0.45\textwidth]{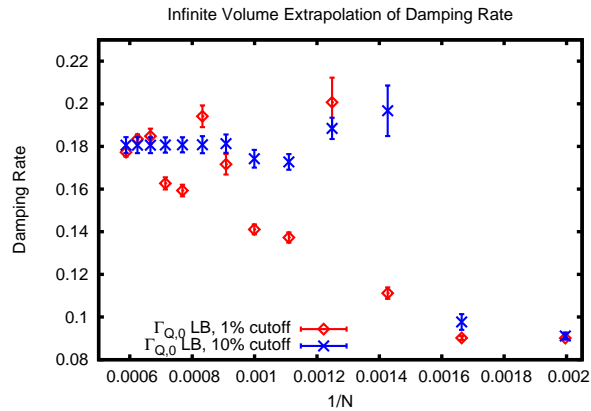}\hfill
\caption{\label{fig:ap5}
Volume dependence of the quadrupole damping rate for a non-ideal Fermi gas in a Gaussian trap for $\tau_R \omega_\perp=0.1$. Shown are results where the damping rate was extracted using data until late times (at which the amplitude reached one percent of the initial amplitude ('1\% cutoff') as well as an extraction discarding the late time date (cutting the data at an amplitude of ten percent ('10\% cutoff')). See text for details.}
\end{figure}

However, in some cases, such for anharmonic traps and non-ideal equations of state studied in the main part of the text, the continuum extrapolation is computationally too demanding for our present resources. In this case, we performed the infinite volume limit for a certain choice of parameters and then used the difference to a simulation at fixed volume as an indicator for the infinite volume trend at other parameter choices. For instance, Fig.\ref{fig:ap5} displays the volume dependence of the quadrupole damping rate for the case of a non-ideal Fermi gas in a Gaussian trap. As can be seen from this figure, the volume dependence of the damping rate is non-monotonic and starts to converge only for very large volumes. We found that the dependence of the extracted damping rate on the volume is considerably smaller if discarding the late-time, low amplitude simulation data. We attribute this to the fact that when the mode amplitude becomes low, numerical noise starts to contaminate the signal. Nevertheless, we find that for extremely large volumes the extracted damping rate from both procedures is equal within the statistical uncertainty. The difference between the extracted damping rate at low volumes ($N=500$) and high volumes ($N=1700$) for the case of $\tau_R \omega=0.1$ defines an infinite volume trend for the extracted data point at low volumes. This trend (characterized as both a direction and magnitude, and assumed to be independent of $\tau_R \omega_\perp$) has been represented as an arrow on the $N=500$ data points for the quadrupole mode shown in Fig.~\ref{fig:GT}. For completeness, we mention that the breathing mode does not suffer from the same severe volume dependence as found for the quadrupole mode. Thus, a standard infinite volume extrapolation is possible in this case.

\end{appendix}

\bibliographystyle{apsrev} \bibliography{sqifs}

\end{document}